\documentclass[%
 aip,
 amsmath,amssymb,
 reprint,%
]{revtex4-1}

\usepackage{graphicx}
\usepackage{dcolumn}
\usepackage{bm}

\usepackage[utf8]{inputenc}
\usepackage[T1]{fontenc}
\usepackage{mathptmx}
\usepackage{etoolbox}

\usepackage[caption=false,lofdepth,lotdepth]{subfig}
\usepackage{bm}

\makeatletter
\def\@email#1#2{%
 \endgroup
 \patchcmd{\titleblock@produce}
  {\frontmatter@RRAPformat}
  {\frontmatter@RRAPformat{\produce@RRAP{*#1\href{mailto:#2}{#2}}}\frontmatter@RRAPformat}
  {}{}
}%
\makeatother
\begin{document}

\preprint{AIP/123-QED}

\title[Superfluid excitations in rotating two-dimensional ring traps]{Superfluid excitations in rotating two-dimensional ring traps}

\author{Guilherme Tomishiyo}

\author{Lucas Madeira}
 \email{madeira@ifsc.usp.br}
\affiliation{ 
Instituto de F\'isica de S\~ao Carlos, Universidade de S\~ao Paulo, CP 369, 13560-970 S\~ao Carlos, Brazil}

\author{M\^onica A.~Caracanhas}
\affiliation{ 
Instituto de F\'isica de S\~ao Carlos, Universidade de S\~ao Paulo, CP 369, 13560-970 S\~ao Carlos, Brazil}

\date{\today}

\begin{abstract}
We studied a rotating Bose-Einstein condensate confined in ring trap configurations that can be produced starting with a bubble trap confinement, approximated by a Mexican hat and shifted harmonic oscillator potentials. Using a variational technique and perturbation theory, we determined the vortex configurations in this system by varying the interparticle interaction and the angular velocity of the atomic cloud. We found that the phase diagram of the system has macrovortex structures for small positive values of the interaction parameter, and the charge of the central vortex increases with rotation. Strengthening the atomic interaction makes the macrovortex unstable, and it decays into multiple singly-charged vortices that arrange themselves in a lattice configuration. We also look for experimentally realizable methods to determine the vortex configuration without relying upon absorption imaging since the structures are not always visible in the latter. More specifically, we study how the vortex distribution affects the collective modes of the condensate by solving the Gross-Pitaevskii equation numerically and by analytical predictions using the sum-rule approach for the frequencies of the modes. These results reveal important signatures to characterize the macrovortices and vortex lattice transitions in the experiments.
\end{abstract}

\maketitle

\section{Introduction}

A defining property of a superfluid is how it responds to an external rotation. Despite the irrotational nature of the superfluid flow, it is understood that a superfluid can rotate, but it does so by nucleating quantized vortices that carry discrete units of vorticity, which reveals the macroscopic quantum nature of the superfluid. When subjected to rotation, a quantum gas might display collective modes or, in some situations, develop one or more quantized vortices: lines of vanishing density around which the phase of the condensate changes by an integer multiple of $2\pi$ \cite{Pitaevskii2016, Pethick2008}.

Early experiments in the field used harmonic potentials to trap the gas \cite{Anderson1995, Davis1995, Bradley1995}. This naturally limits the rotation frequency, as the centrifugal effects weaken the confinement of the trap. For sufficiently large angular velocities, multiple vortices develop in the bulk of the condensate, and the gas becomes populated by a vortex lattice \cite{Butts1999,Chevy2000,Abo-Shaeer2001}, with the average velocity field due to the many vortices mimicking a rigid body velocity field. For even higher velocities, the system is expected to enter a highly degenerate state known as the lowest Landau level \cite{Fetter2008}, with a Hamiltonian mathematically equivalent to the one of an electron moving in a 2D uniform magnetic field, in the context of the quantum Hall effect \cite{Tong2016,Fletcher2021,Mukherjee2022}.

Despite ultracold atoms being usually confined in magnetic or optical traps, typically producing a harmonic profile, increasing interest has been given to Bose-Einstein condensates (BECs) trapped in more exotic potentials. The overlap of magnetic and optical dipole traps allowed the production of a quadratic plus quartic trap radial profile. This enabled to reach arbitrary high angular velocities to excite macrovortices and vortex lattices structures in the superfluid, leading to interesting new configurations which have been studied in several papers~\cite{Fetter2001rotating,Kasamatsu2002,Lundh2002,Kavoulakis2003,Fetter2005}.

In this context of exotic traps, the ring-shaped traps are of particular interest \cite{Salasnich1999,Nugent2003,Aftalion2004,Cozzini2006,Wen2011} as they can be employed as a storage ring for coherent atom waves, enabling the investigation of persistent currents, Josephson effects, phase fluctuations, and high-precision gravitational interferometry. In a ring guide with a repulsive barrier, superfluids can behave analogously to Josephson junctions in superconductors~\cite{Albiez2005,Kwon2020}. Of particular interest is the realization of the atomic analog of a superconducting quantum interference device (SQUID)~\cite{Ramanathan2011,Ryu2013,Wright2013a,Eckel2014a,Ryu2020}. While superconducting SQUIDs are very sensitive magnetometers, their atomic counterparts behave the same concerning rotations. Besides the intrinsic interest in quantum devices, ring traps can also be used to study fundamental questions in statistical physics, for example, the thermalization of a BEC with periodic boundary conditions~\cite{Aidelsburger2017}.
Theoretical work concerning toroidal condensates has investigated its elementary excitations~\cite{Salasnich1999,Nugent2003}, for which the spectrum is intensely affected by the geometry. Also, the vortex dynamics are strongly distorted by such geometry~\cite{Guenther2020} as well as the stability of multiply quantized toroidal currents~\cite{Cozzini2006,Wen2011}, which is also of interest from a fundamental perspective.  Similar to quartic potentials, giant vortex and vortex lattice configurations have also been studied in toroidal traps \cite{Aftalion2004,Cozzini2006,Wen2011}. Aftalion et al.~\cite{Aftalion2004} investigated the case of a quartic minus harmonic potential numerically, showing the 3D configuration of vortices established for different external rotation values.

Recently, a very rich potential under theoretical and experimental scrutiny is the bubble trap potential, proposed by Garraway and Zobay in 2001~\cite{Zobay2001,Zobay2004}. The bubble trap is an adiabatic potential formed by superpositioning a static inhomogeneous magnetic field with a spatially homogeneous radio frequency (RF). This confines the atoms to an isomagnetic surface. In typical experiments, gravity cannot be neglected, and the resulting equilibrium configuration pools the atoms at the bottom of the surface \cite{Colombe2004, Garraway2016,Arazo2021}. To overcome this difficulty, ongoing experimental efforts exist to realize the bubble trap in microgravity conditions aboard the Cold Atom Laboratory in the International Space Station~\cite{Lundblad2019,Aveline2020,Carollo2022,Lundblad2023}. In the future, the experiment is expected to include the possibility of rotations in the system.

Besides providing stronger than harmonic confinement for the condensate, the bubble trap also allows for control over the topology of the system. The radius of the bubble can be controlled by tunning the magnetic gradients and the RF frequency \cite{Goer2021}. By varying the parameters of the field, one can control how strongly the atoms are confined to the isomagnetic surface. This allows for studying nearly two-dimensional (2D) gases bound to surfaces \cite{Perrin2017}. The topology has non-trivial consequences over many properties of the gas. We mention, for instance, the effects of the dimensionality over the condensation temperature, which are already notorious \cite{Pitaevskii2016, Pethick2008}. Padavic and Sun analyzed how the collective modes of a spherical condensate change in the transition from a filled to a hollow sphere, finding a dip in the breathing modes and a restructuring in the surface modes due to the emergence of a second interior surface \cite{Padavic2017, Sun2018}.

Motivated by ring trap configurations that can be produced starting with a bubble trap confinement~\cite{Garraway2016,Goer2021}, we want to study two approximations to these trapping potentials: the Mexican hat (MH) and shifted harmonic oscillator (SHO). Using a variational technique and numerical methods, we construct a phase diagram of the arrangement of the vortices as the interaction parameter and the angular velocity of the gas vary. We make predictions for discontinuous transitions, where the total vortex charge increases, and continuous transitions, where only the charge distribution varies. For weak interactions and low rotation speeds, the system is characterized by a central macrovortex, but as interaction increases, it breaks into multiple charge $1$ vortices. The shifted harmonic oscillator potential can withstand a broader range of interaction parameters before breaking the macrovortex compared to the Mexican hat potential. We also develop sum rules and numerical predictions using real and imaginary time simulations of the Gross–Pitaevskii equation (GPE) for the monopole and quadrupole collective modes of the gas, searching for a signature of the charge and charge distribution of vortices. We find that monopole frequencies increase with rotation only when there is a corresponding increase in vortex charge, and it is larger for a vortex lattice than for a macrovortex. At the same time, there is a frequency split dependent on the total angular momentum between quadrupole modes co- and anti-rotating with the gas.

This work is structured as follows: in Sec.~\ref{sec:Formalism}, we present the bubble trap, and we analyze two limits of it -- the Mexican hat and the shifted harmonic oscillator. Then, in Sec.~\ref{sec:ringConfiguration}, we present the total Hamiltonian of the system, including rotation and strong confinement in the rotational direction, which gives the ring configuration. The energy of the system is minimized using a wave function described in Sec.~\ref{sec:diagram}, where we also present our results for the MH and SHO potentials. Seeking experimental alternatives to characterize the vortex configurations of the gas, in Sec.~\ref{sec: COllective modes} we introduce the sum rules formalism for evaluating the monopole and two-dimensional quadrupole collective modes, with special attention to how the vortex configuration affects these frequencies. Finally, we present our conclusions in Sec.~\ref{sec:conclusions}. In the Appendices, we provide more details on the derivation of results reported in the main text. Appendix~\ref{sec:app_sum_rules} deals with the collective modes and sum rules. We define the relevant excitation operators, present the two- and four-modes approaches to the quadrupole mode, calculate the necessary moments, and discuss the virial theorem. The compressibility sum rules are explained in Appendix~\ref{compressibility}.

\section{Bubble trap with rotation} 
\subsection{Bubble trap}
\label{sec:Formalism}

Experimentally, a bubble trap can be constructed by combining a linearly or circularly polarized radio frequency (RF) and a spatially inhomogeneous static magnetic field. In this setting, the atoms are said to be ``dressed'' by the RF~\cite{Zobay2001,Zobay2004}, and are trapped close to an isomagnetic surface.

In particular, the isomagnetic surface of an ellipsoid of revolution corresponds to the magnetic resonance, which occurs when the RF photon energy matches the Zeeman splitting given by the original static magnetic trap. Under adiabatic conditions, i.e., if atoms cross the region with small enough velocities, it can be shown that they will feel an adiabatic potential given by~\cite{Zobay2001,Garraway2016,Perrin2017}
\begin{equation}
\label{ellipse}
\tilde{V}_{\textrm{Bubble}}(r) = M_F \sqrt{\left(\frac{1}{2}\frac{\tilde{M}\tilde{\omega}_0^2 \tilde{r}^2}{2}-\hbar\tilde{\Delta}\right)^2+(\hbar \tilde{\Theta})^2},
\end{equation}
where $\tilde{\Delta}$ corresponds to the detuning between the applied RF field and the energy states, and $\tilde{\Theta}$ is the Rabi coupling between these states. In the limit $\tilde{\Delta} = \tilde{\Theta}=0$, the potential of Eq.~(\ref{ellipse}) reduces to a usual harmonic oscillator trap. On the other hand, for large $\tilde{\Delta}$, it can be approximated near its minimum by a radially shifted harmonic trap with frequency $\tilde{\omega}_{\rm SHO}=  ({M_F \tilde{\Delta}}/{\tilde{\Theta}})^{1/2}\tilde{\omega}_0$.

Hereafter we reserve the tilde notation for quantities with dimensions. Dimensionless quantities are constructed considering
$r\equiv\tilde{r}/{\tilde{\ell}_0}$ (with $\tilde{\ell}_0\equiv[{\hbar}/({\tilde{M}\tilde{\omega}_0)}]^{1/2}$), $\Theta\equiv\tilde{\Theta}/{\tilde{\omega}_0}$, $\Delta\equiv\tilde{\Delta}/{\tilde{\omega}_0}$, and energies in units of $\hbar\tilde{\omega}_0$.
Choosing the atomic hyperfine state $M_F=2$ yields
\begin{equation}
\label{ellipsedimensionless}
V_{\textrm{Bubble}}(r) =2 \sqrt{\left(\frac{r^2}{4}-\Delta\right)^2+\Theta^2},
\end{equation}
with a trap minimum corresponding to $r_{\rm min}= 2\sqrt{\Delta}$.

We considered two limits of the bubble trap potential. For that, it is convenient to rewrite Eq.~(\ref{ellipsedimensionless}) in terms of $r_{\rm min}$,
\begin{equation}
\label{secondform}
{V}_{\textrm{Bubble}}({r}) = \frac{r_{\rm min}^2}{2} \sqrt{\left(\frac{r^2}{r_{\rm min}^2}-1\right)^2+\sigma^2},
\end{equation}
where we defined the width $\sigma \equiv \Theta/\Delta $.

\begin{figure}[!htb]
    \centering
    \begin{minipage}{0.5\textwidth}
        \centering
        \subfloat[\label{figSHOa}]{\includegraphics[width=1.0\textwidth]{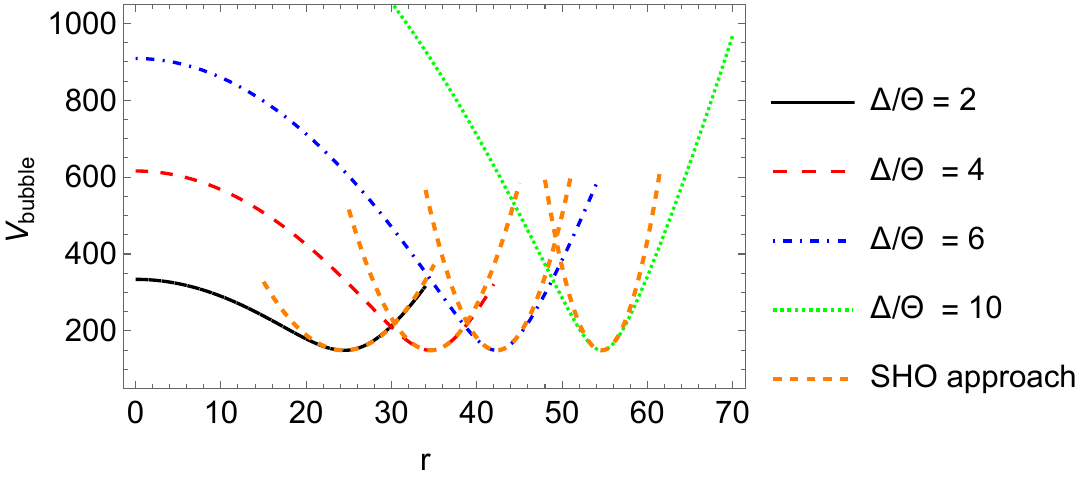}}
    \end{minipage}\hfill
    \begin{minipage}{0.5\textwidth}
        \centering
        \subfloat[\label{figSHOb}]{\includegraphics[width=1.0\textwidth]{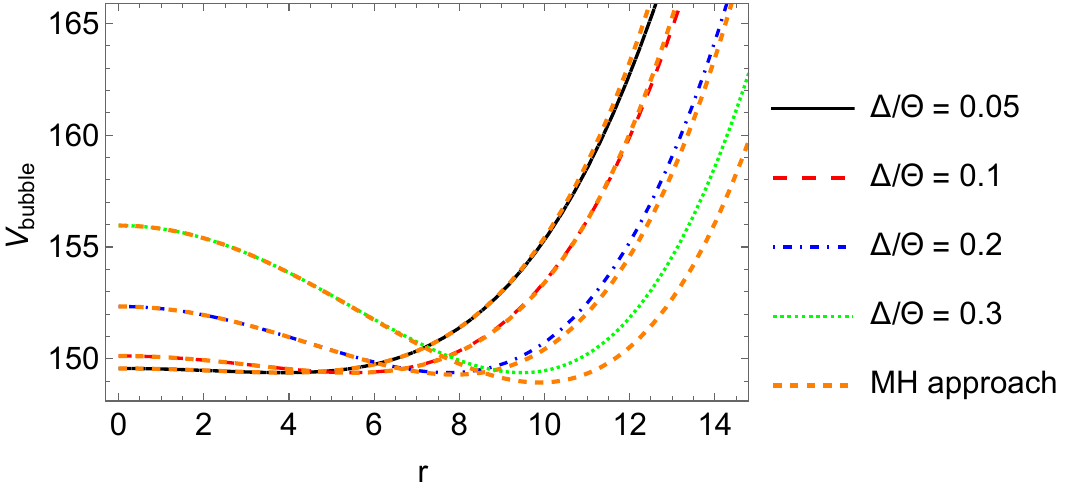}}
    \end{minipage}
\caption{Comparison between bubble trap potentials, given by Eq.~(\ref{ellipsedimensionless}), with $\Theta=75$ and several values of $\Delta$. We employed two approximations: (a) the shifted harmonic oscillator, Eq.~(\ref{SHO_potentialapp}), and (b) the Mexican hat, Eq.~(\ref{MH_potentialapp}). The latter is valid for small values of the detuning parameter ($\Delta/\Theta\ll 1$).
\label{figSHO}}
\end{figure}

\textit{Shifted harmonic oscillator.---}
Considering the expansion of Eq.~(\ref{secondform}) around the trap minimum, $r=r_{\rm min}+\delta r$, we obtain 
\begin{equation}
\label{SHO_potentialapp}
{V}_{\textrm{Bubble}}({r}) = \frac{r_{\rm min}^2\sigma}{2} + \frac{\delta r^2}{\sigma} + O(\delta r^3).
\end{equation}
This allows us to characterize the bubble trap as a shifted harmonic oscillator potential with trap frequency $\omega_{\rm SHO}=\sqrt{2/\sigma}=\sqrt{2{\Delta}/{\Theta}}$. From the expression above, we see that, for a fixed value of $\Theta$, as we increase the value of the detuning parameter $\Delta$, we get a thinner shell potential with a larger radius for $\Delta > \Theta$, as depicted in Fig.~\ref{figSHOa}.

\textit{Mexican hat.---}
We explored another limit of Eq.~(\ref{secondform}) considering the expansion of $r\rightarrow 0$,
\begin{equation}
\begin{split}
{V}_{\textrm{Bubble}}(r) =&  \frac{r_{\rm min}^2 \sqrt{1+\sigma^2} }{2}-\frac{1}{2\sqrt{1+\sigma^2}} r^2 \\
+&\frac{\sigma^2}{4 r_{\rm min}^2 (1+\sigma^2)^{3/2}}r^4+O(r^6).  
\end{split} \label{MH_potentialapp}
\end{equation}
As illustrated in Fig.~\ref{figSHOb}, for the Mexican hat potential, we need to be in the regime of small radii and large $\sigma$ values ($\sigma \gg 1$, a thick bubble). In contrast, the SHO agreement occurs in the opposite regime, with large radii and small $\sigma$ values ($\sigma < 1$, a thin bubble).

Further aspects of the validity of the two limiting approximations for the actual ring trap are discussed in Appendix~\ref{app:approximations}, where we compare the numerically calculated wave function of the bubble trap with that of the approximations, the MH and SHO potentials.

\subsection{Ring trap potentials and the rotating frame Hamiltonian}
\label{sec:ringConfiguration}

In the previous section, we were treating a spherically-shaped bubble trap. However, this work aims to explore ring trap configurations under rotation. A ring trap can be produced by superposing a bubble trap with a vertical optical trap. That gives us a widely tunable trap, with the radius of the ring varying from $10\mu$m to $100 \mu$m, and with a fixed coordinate $z_0$ in the vertical direction for the trapped atoms \cite{Goer2021}. It is straightforward to derive the SHO and MH potential profiles in 2D from Eqs.~(\ref{SHO_potentialapp}) and (\ref{MH_potentialapp}), respectively, after the replacement $r=\sqrt{\rho^2+z_0^2}$, where $\rho$ is the usual polar coordinate. Figure~\ref{figSHO} and the previous discussions still hold if we take $z_0=0$ and substitute $r$ by $\rho$. For the SHO case, we have to consider additionally $\rho \gg z_0$, which is a very reasonable experimental condition \cite{Garraway2016,Goer2021}.

An alternative way of generating a ring configuration is dynamically, with the gas initially pooled at the bottom of a bubble trap due to the action of gravity. Then, suppose we increase the rotation rate. In that case, centrifugal effects propel the gas over the isomagnetic surface. The disposition of the gas is then one of a dynamical ring. However, the precise parameterization of the radius of the ring and the minimum height of the potential as a function of the angular speed is a challenging problem \cite{Guo2020}.

There are two main experimental methods for rotating the atoms. One consists in deforming the trap and rotating the deformation \cite{Jin1996, Mewes1996}. The other is based on a blue-detuned laser to stir the atoms in the ring trap \cite{Marzlin1997, Dum1998}. The rotating frame Hamiltonian $H'$ is related to the one in the laboratory frame $H$ by
\begin{eqnarray}\label{rotFrame} \nonumber 
H^{'}   &=& H - {\bm \Omega}\cdot\bm{L}, \\ 
H^{'}   &=&\sum_{i=1}^{N}\left[\frac{\bm{p}_i^2}{2}+V(\bm{r}_i)\right]+\frac{G_{2D}}{N} \sum_{i<j}  \delta(\bm{r}_i-\bm{r}_j) \nonumber \\
        &-&\sum_{i=1}^{N}\bm{\Omega}\cdot\left(\bm{r}_i\times\bm{p}_i\right),
\end{eqnarray}
where the usual polar coordinates give the two-dimensional position of the atoms, $\bm{r}_i\equiv (\rho_i,\varphi_i)$, and $N$ is the total number of particles. In Eq.~(\ref{rotFrame}) we also introduced the angular velocity $\bm{\Omega}$ and the dimensionless 2D coupling constant $G_{2D}\equiv N \tilde{g}_{2D}/ (\hbar\tilde{\omega}_0 \tilde{\ell}_0^2)$, where $\tilde{g}_{2D}$ characterizes the inter-atomic interactions. The 2D interaction strength is related to its three-dimensional (3D) counterpart by 
\begin{equation}
\tilde{g}_{2D}=\frac{\tilde{g}_{3D}}{\sqrt{2\pi} \tilde{Z}},
\end{equation}
where $\tilde{Z}$ is associated with the length scale of the optical-trap confinement in the $z$-direction.

\section{Phase Diagram}
\label{sec:diagram}

\begin{figure*}[!htb] 
\centering
\subfloat[\label{fig:MexPhase1a}]{\includegraphics[width=0.4\textwidth]{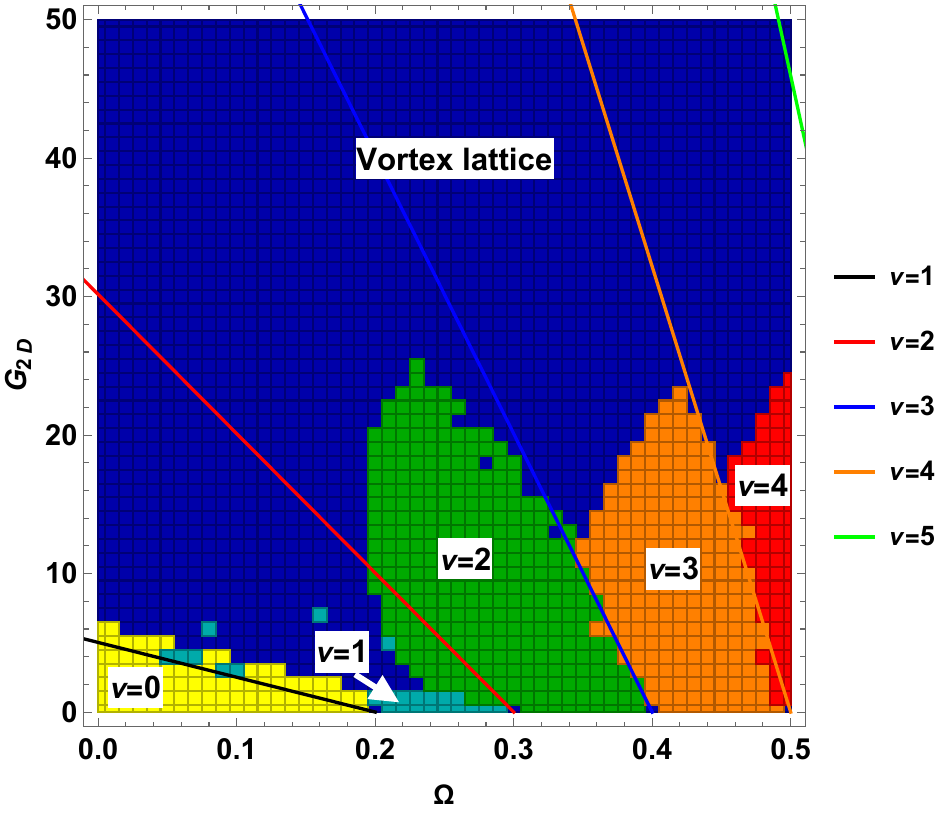}}
\subfloat[\label{fig:MexPhase1b}]{\includegraphics[width=0.125\textwidth]{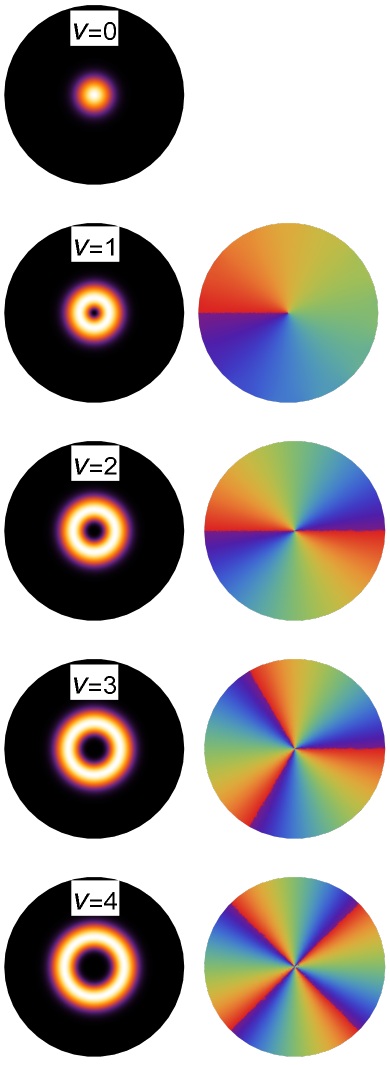}}
\subfloat[\label{fig:MexPhase2a}]{\includegraphics[width=0.4\textwidth]{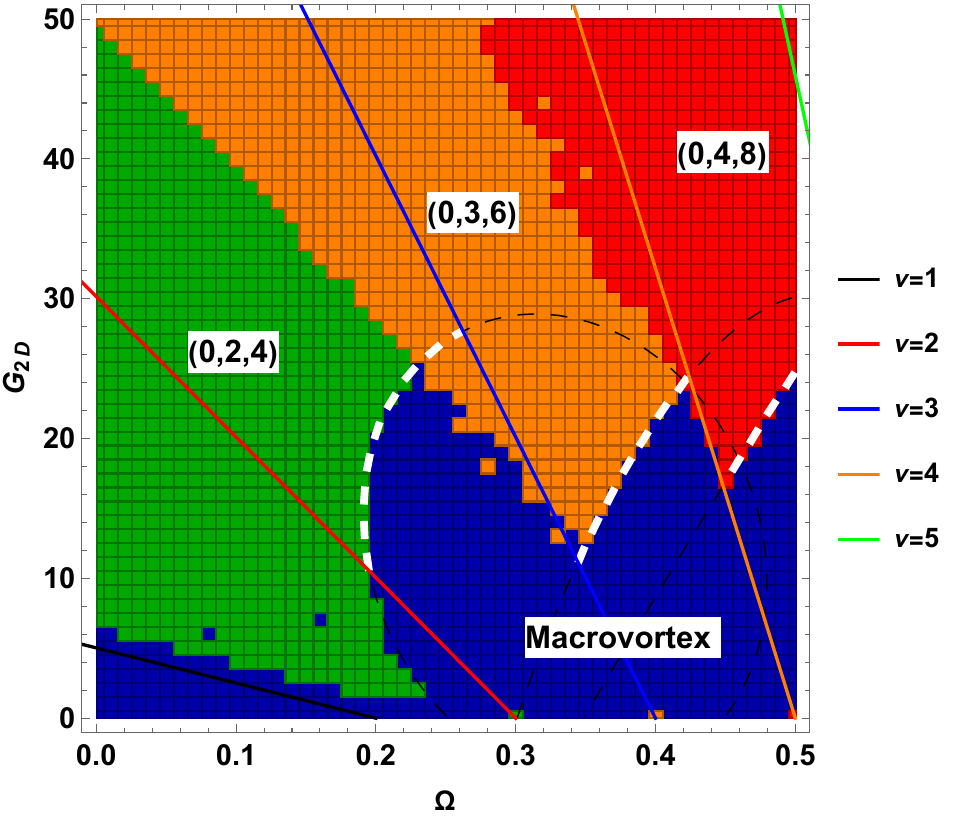}}
\subfloat[\label{fig:MexPhase2b}]{\includegraphics[width=0.11\textwidth]{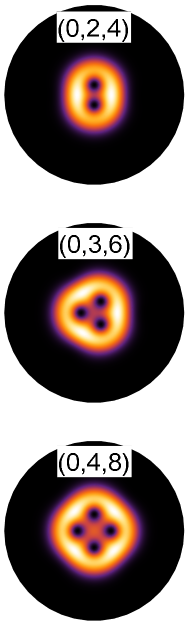}}
\caption{Phase diagram of the Mexican hat potential, Eq.~(\ref{eq:MHpotential}), with $\lambda=0.1$. The parameter space corresponds to $0.01\leqslant \Omega \leqslant 0.50$ and $0\leqslant G_{2D} \leqslant 50$. For clarity, we decomposed the phase diagram into two figures, panels (a) and (c), where the former deals with the macrovortices and the latter with the vortex lattices. (a) The colors represent the region of each macrovortex configuration, i.e., the pure states, with increasing values of vorticity (starting from the left side, with $\nu$ going from $0$ to $4$) as predicted by variational calculations. The lines represent phase transitions calculated with perturbation theory, Eq.~(\ref{OmegaCrit}).
(b) Phase and density profiles of the macrovortex states. The first profile shows a zero vorticity configuration represented in the phase diagram by the area below the solid black line. (c) The blue region collects the zero vorticity regime plus the macrovortex regions, whereas the other colored regions correspond to different vortex lattices indexed by $(0,m_0,2m_0)$. The solid lines are the same as panel (a). The dashed curves delimit the stability region of each macrovortex configuration, with the highlighted white thick dashed curves representing the transition from a macrovortex to a vortex lattice phase obtained with the stability analysis. Above (below) the white curves, the system is expected to be in a lattice (macrovortex) configuration. (d) Vortex lattice density profiles.}
     \label{fig:MexPhase1}
\end{figure*} 

To characterize the phase diagram of the system, we applied the variational technique as described in the following. Provided the interaction parameter $G_{2D}$ is sufficiently small, we can consider the eigenstates of the harmonic potential without radial excitations, and angular momentum $m$ as our unperturbed states $|\Phi_m \rangle$ \cite{Jackson2004b}. As in Ref.~\cite{Jackson2004a}, we used a linear combination of the low-energy angular momentum eigenstates of the harmonic oscillator potential to represent the order parameter. The authors were interested in the traditional quadratic plus quartic confinement, while in this work, we employ this expansion for the MH and SHO traps. This complete basis is justified by the weak interaction regime that we are considering \cite{Butts1999}. 
Therefore, the condensate wave function can be represented by the variational ansatz
\begin{equation} \label{ansatzMH}
|\psi\rangle=\sum_{m=0}^{N_{\rm max}} c_m |\Phi_m \rangle,
\end{equation} with
\begin{equation} \label{ansatzgauss}
\langle\boldsymbol{r}|\Phi_m\rangle= \frac{\rho^m e^{i m \varphi}}{\sqrt{\pi m!}}e^{-\frac{\rho^2}{2}},
\end{equation} and 
\begin{equation} \label{eq:normCondition}
\sum_{m=0}^{N_{\rm max}} |c_m|^2=1.
\end{equation}
These functions are the basis of the functional space over which we minimize the energy.

It should be mentioned that the radial derivative of the total wave function, Eq.~(\ref{ansatzMH}), must vanish at $\rho=0$. The orbitals in our ansatz, Eq.~(\ref{ansatzgauss}), always satisfy this requirement for $m\neq 1$. Hence, if $c_1\neq 0$, the density must vanish at the center, which happens if $c_0=0$. In other words, the issue with the derivative of the wave function appears only if $c_0$ and $c_1$ are both non-zero. All results in the manuscript obey this requirement; hence, the wave function of Eq.~(\ref{ansatzMH}) has the correct boundary conditions.

We begin by choosing a value for the rotation speed $\Omega$ and the interaction parameter $G_{2D}$. We find numerically the set of coefficients which minimize the energy, given by the expected value of Eq.~(\ref{rotFrame}). It is possible to classify this $(\Omega,G_{2D})$ pair as a macrovortex if $|c_m|^2\approx1$ (within a $10^{-3}$ tolerance) for only one $m$ or as a vortex lattice if there are multiple coefficients contributing to the wave function. Then, the procedure is repeated for several values of $(\Omega,G_{2D})$.

\subsection{Mexican hat}
\label{sec:mh_phase_diagram}

\subsubsection{First-order perturbation theory}
\label{sec:MH_first_order}
For the Mexican hat, it is convenient to rewrite Eq.~(\ref{MH_potentialapp}) as
\begin{equation} \label{eq:MHpotential}
V(\rho) = \frac{1}{2}(-\rho^2+\lambda \rho^4),
\end{equation}
where we rescaled the distances with $\rho=r/(1+\sigma^2)^{1/4}$, and we introduced the parameter $\lambda$,
\begin{equation}
\lambda = \frac{\sigma^2}{2 r_{\rm min}^2 (1+\sigma^2)^{1/2}}, 
\end{equation}
which depends explicitly on $\sigma$ and $r_{\rm min}$. Equation~(\ref{eq:MHpotential}) is the trap potential entering into Eq.~(\ref{rotFrame}).

To use the eigenfunctions given by Eq.~(\ref{ansatzgauss}) we also require $\lambda \ll 1$. 
In Fig.~\ref{fig:MexPhase1}, we present our results for $\lambda=0.1$. Hereafter we denote the macrovortex charge by $\nu$. In the parameter space explored in this work, $0.01\leqslant \Omega \leqslant 0.50$ and $0\leqslant G_{2D} \leqslant 50$, we observed macrovortices with charge varying from $\nu=0$ up to 4, see Fig.~\ref{fig:MexPhase1a}. The density profiles and respective phases of the wave functions in each of these regions are illustrated in Fig.~\ref{fig:MexPhase1b}. In addition, we observed an upper (blue) region not described by one single coefficient $c_m$, signalling the appearance of vortex lattices.

It is worth noting the results reported in Ref.~\cite{Jackson2004a} for $\Omega >1$ are similar to ours for $\Omega <1$, Fig.~\ref{fig:MexPhase1}. Reference~\cite{Jackson2004a} uses a different trap, the conventional harmonic plus quartic potential, and the same basis for the order parameter. The reason is that, since the centrifugal term inverts the global sign of the quadratic term of the potential, they obtain an effective MH potential.

A few words about the parameter space we chose to employ in this work are in order. The confinement in the $z$-direction must be strong enough so that the relevant dynamics takes place only in the $xy$-plane. A length scale $Z=0.1$ is one order of magnitude smaller than the harmonic oscillator length, thus accomplishing this goal. The interaction strength is given by $G_{2D}= \sqrt{8\pi} N {a}_s/Z$. Thus, even for the maximum value of $G_{2D}=50$ that we considered, we have $N a_s\sim 1$. Hence, we are far from the Thomas-Fermi regime, and the ansatz of Eq.~(\ref{ansatzMH}) is valid. Concerning the rotation frequency, we use values that stay below the quadratic trap frequency, that is, $\Omega <1$.

Next, we wanted to compare our numerical variational results with perturbation theory predictions. We calculated the expectation value of $H^{'}$ in the pure macrovortex states $|\Phi_\nu\rangle$ by using first-order perturbation theory to consider the interaction energy per particle and the anharmonicity contribution of the trap \cite{Jackson2004a}. Then we compared the energy of states $\nu$ and $\nu+1$ to extract the macrovortex phase boundaries. This gives us an analytical expression for the critical frequency $\Omega$, which delimits the transition between macrovortices states in the $G_{2D}\times \Omega$ phase diagram.

The energy $E_\nu$ of a pure macrovortex state $|\Phi_\nu\rangle$ is given by
\begin{eqnarray}
E_\nu  &=& \langle \Phi_\nu|H^{'}|\Phi_\nu \rangle\nonumber \\
     &=&\frac{1}{2}\lambda \; (1+\nu)(2+\nu)+\frac{G_{2D}}{4\pi}\frac{(2\nu)!}{2^{2\nu}(\nu!)^2} \nonumber \\
     &-&\nu\Omega.
\end{eqnarray}
Comparing the energies $E_\nu$ and $E_{\nu+1}$, we obtain the critical frequency for the macrovortex phase transition,
\begin{equation} \label{OmegaCrit}
\Omega\left(G_{2D}\right) = \lambda \; (1+\nu) - \frac{G_{2D}}{2\pi}\frac{(2\nu-2)!}{2^{2\nu}(\nu-1)!\nu!}.
\end{equation} 
Equation~(\ref{OmegaCrit}) is a line in the $G_{2D}\times \Omega$ phase diagram, with the intercept proportional to the anharmonicity.

We included the first-order perturbation theory results, Eq.~(\ref{OmegaCrit}), as solid lines in Fig.~\ref{fig:MexPhase1a}. The lines delimit reasonably well the transition between the different macrovortex configurations for small values of $G_{2D}$. However, as we increase the value of the interaction, perturbation theory no longer accurately predicts the contours of transitions.


\subsubsection{Stability analysis}
\label{sec:second_order_mh}

Let us consider the state of the system, starting from a macrovortex configuration, as we increase the interaction parameter $G_{2D}$. Initially, the state is characterized by some coefficient $|c_{m_{0}}|^2=1$ signalling unit occupation of the $m_0$ component in Eq.~(\ref{ansatzMH}). As $G_{2D}$ increases, the occupation of $m_0$ slowly decreases and other states start to contribute to the total wave function. Eventually, the occupations of these states cease to be negligible and the system transitions from a pure macrovortex to a vortex lattice configuration.

\begin{figure*}[!htb] 
    \centering
    \begin{minipage}{0.33\textwidth}
        \centering
        \subfloat[\label{fig:024_mh}]{\includegraphics[width=0.9\textwidth]{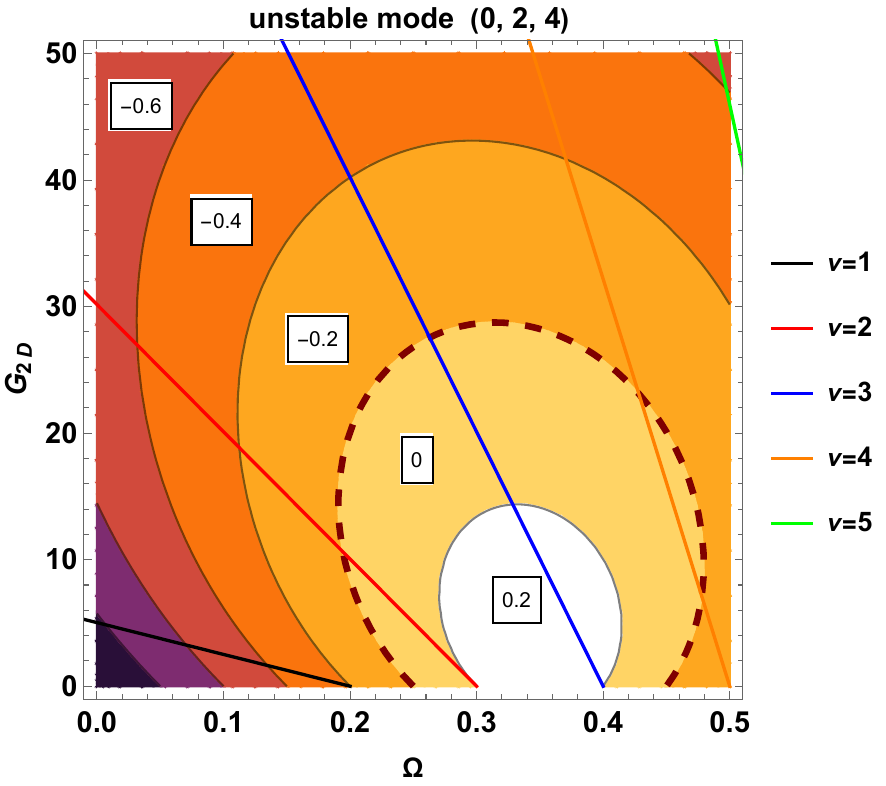}}
    \end{minipage}\hfill
    \begin{minipage}{0.33\textwidth}
        \centering
        \subfloat[\label{fig:036_mh}]{\includegraphics[width=0.9\textwidth]{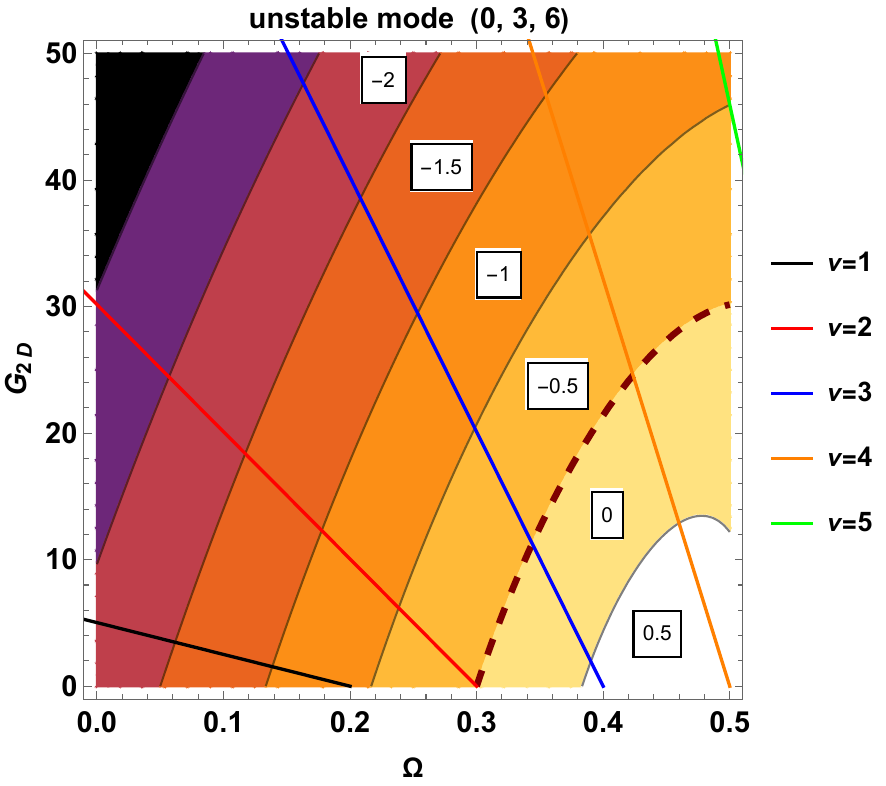}}
    \end{minipage}\hfill
    \begin{minipage}{0.33\textwidth}
        \centering
        \subfloat[\label{fig:048_mh}]{\includegraphics[width=0.9\textwidth]{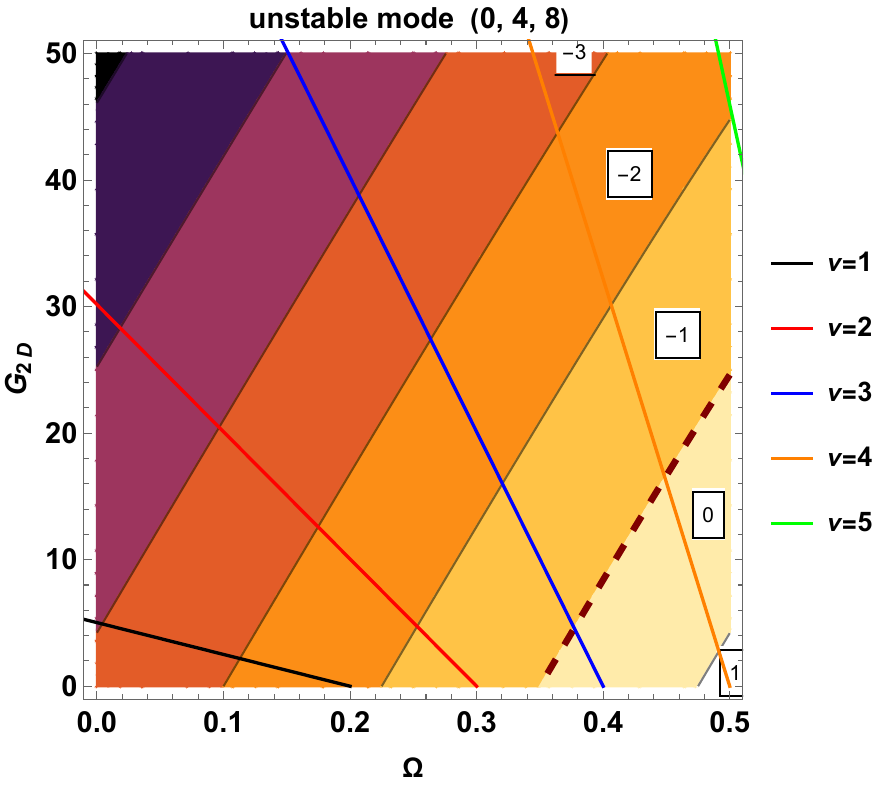}}
    \end{minipage}\hfill
      \caption{The figures show the value of the smallest eigenvalue of the stability matrix for the pure states $m_0 =2,3,4$. The boundary of the positive region in each diagram corresponds to a phase transition from a macrovortex to vortex lattice regime. The solid lines are the results of first-order perturbation theory calculation, Eq.~(\ref{OmegaCrit}).}
      \label{fig:Mex2ndOrder}
\end{figure*}

Using the same variational results reported in the previous section, we now focus on the macrovortex to vortex lattice transitions. In Fig.~\ref{fig:MexPhase2a} we indicate all the individual macrovortex regions ($\nu=0$ to 4) in blue. The other areas in the phase diagram correspond to vortex lattices. The classification of each of the lattice regions is given by a triple of numbers $(m_1,m_0,m_2)$, which label the most occupied states, with $|c_{m_0}|^2+|c_{m_1}|^2+|c_{m_2}|^2 \approx 1$ (within a $10^{-3}$ tolerance) and $|c_{m_0}|^2>|c_ {m_1}|^2,|c_{m_2}|^2$. For completeness, we keep the solid lines corresponding to Eq.~(\ref{OmegaCrit}) that would represent the boundaries of the macrovortex regions. In Fig.~\ref{fig:MexPhase2b}, we show the density profiles of the vortex lattices.

The transition from the macrovortex state to the lattice configuration formed by singly-charged vortices occurs for sufficiently high interaction parameter values, which leave the macrovortex configuration energetically unstable. Analyzing the coefficients $c_m$, we also conclude that the most likely lattice mode for the decay of the macrovortex has the $(0,m_0,2m_0)$ combination of coefficients. This indicates that this mode is the most unstable one, which we confirmed by performing a stability analysis which we now describe.

We constructed a stability matrix formed by the derivative of the energy with respect to the coefficients of the system close to the transition \cite{Jackson2004a}. Suppose the system is stable in a macrovortex configuration characterized by angular momentum $m_0$. In that case, the energy derivative with respect to all $c_m$ must vanish, and the eigenvalues must be positive. The total energy is given by
\begin{eqnarray} \label{TotalEnergy} \nonumber
E  &&= \langle \psi(\bm{r})|H^{'}|\psi(\bm{r})\rangle \\ 
&&= \sum_m |c_m|^2 \epsilon_m \nonumber \\
&& +\frac{1}{S}\sum_{m,n,l,k} c_m^* c_n^* c_lc_k \langle m, n| V_{\rm int} | l,k\rangle \; \delta_{m+n,l+k},
\end{eqnarray} where $\epsilon_m \equiv \lambda (m+1) (m+2)/2- m \Omega$ corresponds to the energy of a particle in the absence of interaction, $S\equiv\sum_{m} |c_m|^2$, and $V_{\rm int}\equiv (G_{2D}/N) \sum_{i<j}^{N}  \delta(\bm{r}_i-\bm{r}_j)$ is the interaction term. Its matrix elements are
\begin{eqnarray}
 &&\langle m, n| V_{\rm int} | l,k\rangle \nonumber\\ \nonumber &&=  \int d^2\boldsymbol{r}  \int d^2\boldsymbol{r'} \Phi_m^*(\boldsymbol{r})  \Phi_n^*(\boldsymbol{r'}) V_{\rm int}(\boldsymbol{r}-\boldsymbol{r'})\Phi_l(\boldsymbol{r'})  \Phi_k(\boldsymbol{r}) \\ &&=\frac{G_{2D}}{4\pi}\frac{(m+n)!}{2^{m+n}\sqrt{m!n!l!k!}} .  
\end{eqnarray}
Note that $\partial E/\partial c_m = 0$ when $c_{m_0}\to1$ and $c_m\to0$ for $m\neq m_0$, as required.

Since the coefficients are small, we need only to keep terms that are at most bilinear in $c_m$ for $m\neq 0$ in the interaction term. Because of the Kronecker delta, only the terms proportional to $\langle m_i m_0| V_{\rm int} | m_i m_0\rangle$ and $\langle m_0 m_0| V_{\rm int} | m_1 m_2\rangle$ need to be retained in the sum, the latter being the only off-diagonal element in the matrix.

The $m_1$ and $m_2$ numbers characterize the two relevant states at the transition, and the elements of the stability matrix corresponding to the state $m_0$ are related to the second derivative of the energy with respect to $m_1$ and $m_2$,
\begin{equation} \label{eq:mixedDerivative}
\frac{\partial^2 E^{(2)}}{\partial{c_{m_1}}\partial{c_{m_2}}} = 4 c_0^2  \langle m_1 m_2| V_{\rm int} | m_0 m_0\rangle \; \delta_{m_1+m_2,2 m_0}, 
\end{equation} and 
\begin{eqnarray} \nonumber \label{approx1}
\frac{\partial^2 E^{(2)}}{\partial^2{c_{m_i}}} &=& 2 \epsilon_{m_i} + 8 c_0^2 \langle m_i m_0| V_{\rm int} | m_i m_0\rangle + \\
&+&  4 c_0^2  \langle m_0 m_0| V_{\rm int} | m_0 m_0\rangle \nonumber \\
&=& 2 (\epsilon_{m_i}-\epsilon_0) + 8 c_0^2 \langle m_i m_0| V_{\rm int} | m_i m_0\rangle \nonumber \\
&-& 4 c_0^2  \langle m_0 m_0| V_{\rm int} | m_0 m_0\rangle.
\end{eqnarray}
In Eq.~(\ref{approx1}) we used the identity $8 c_0^2  \langle m_0 m_0| V_{\rm int} | m_0 m_0\rangle = - 2 \epsilon_0 $, which can be derived from the fact that $c_m=c_{m_0}$ is an energy minimum (${\partial E}/{\partial c_{m_0}}=0$). After choosing a value for the pure state $m_0$ and two values $m_1$ and $m_2$, we construct a $2\times2$ matrix, with the diagonal values given by Eq.~(\ref{approx1}) and the off-diagonal values given by Eq.~(\ref{eq:mixedDerivative}). This matrix is then diagonalized, and the resulting eigenvalues are used to analyze the stability of the pure state $m_0$. One of the eigenvalues is always positive, making the smallest eigenvalue the relevant one to investigate the stability.

The instability of the pure mode $m_0$ towards the decay to the lattice mode $(m_1,m_0,m_2)$ is illustrated in Fig.~\ref{fig:Mex2ndOrder} for the different types of lattices observed in the variational results, namely (0,2,4), (0,3,6), and (0,4,8). The figures represent the smallest eigenvalue of the stability matrix of the state $m_0$. The boundaries in the $G_{2D}\times \Omega$ diagram where this eigenvalue becomes negative represent the transition to a lattice configuration, which we highlight with a dashed curve.

Although the contour where the smallest eigenvalue changes signs intersects regions with different macrovortex charges $\nu$, a transition to a vortex lattice is only physically sensible if $\nu=m_0$. For example, in Fig.~\ref{fig:024_mh}, where $m_0=2$, it is the portion of the dashed curve between the lines corresponding to $\nu=2$ (red) and $\nu=3$ (blue). Analogously, the same is valid for $\nu=3$ and $\nu=4$ (orange) in Fig.~\ref{fig:036_mh}, and $\nu=4$ and $\nu=5$ (green) in Fig.~\ref{fig:048_mh}.

Furthermore, the instability to decay to the lattice mode is more significant as the negative eigenvalue is larger in magnitude. This explains the competition between lattice modes, which defines the boundaries of the colored regions in the diagram of Fig.~\ref{fig:MexPhase2a}.

To make comparisons easier, we included the dashed curves of Fig.~\ref{fig:Mex2ndOrder} in the $G_{2D}\times \Omega$ phase diagram, Fig.~\ref{fig:MexPhase2a}. The regions where a macrovortex to vortex lattice transition is expected, as discussed above, are highlighted as white dashed curves. We see they reasonably delimit the beginning of the vortex lattice regions obtained with the variational technique.

\begin{figure}[!htb] 
      \centering
    \begin{minipage}{0.35\textwidth}
         \centering
        \subfloat[\label{fig:erot_mh}]{\includegraphics[width=\textwidth]{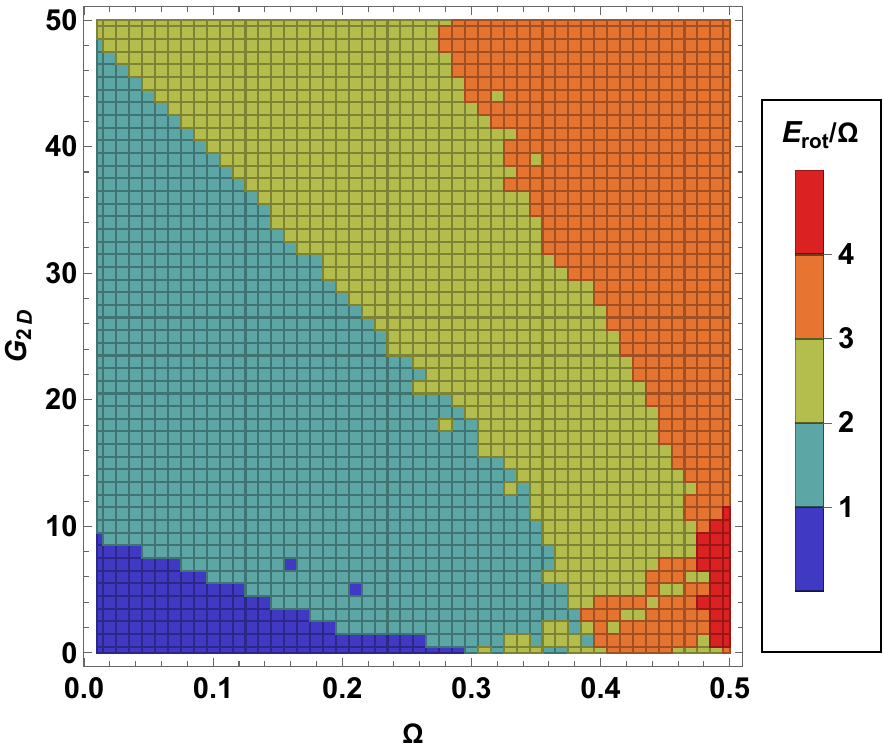}}
    \end{minipage}\hfill
    \begin{minipage}{0.35\textwidth}
        \centering
        \subfloat[\label{fig:ekin_mh}]{\includegraphics[width=\textwidth]{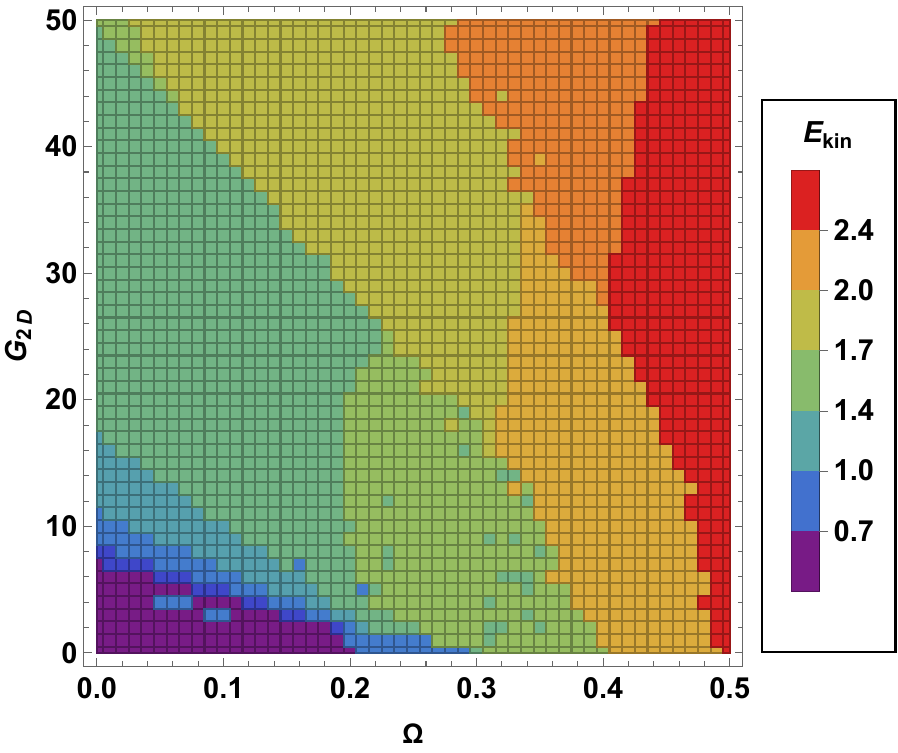}}
    \end{minipage}
      \caption{
Partial energy diagrams for the Mexican hat potential, Eq.~(\ref{eq:MHpotential}), with $\lambda=0.1$, where signatures of the transitions computed with first-order perturbation theory and via stability analysis can be seen. (a) The rotational energy shows an agreement with the perturbation theory calculations, corresponding to transitions between macrovortices with different charges. (b) The kinetic energy signalizes the macrovortex to vortex lattice transitions, as seen in the stability analysis.
}
      \label{fig:MexEnergies}
\end{figure}

The wave function, Eq.~(\ref{ansatzMH}), can also be used to compute the partial energies of the system,
\begin{eqnarray}
\label{eq:partial_energies}
E_{kin} &=& -\frac{1}{2} \int d^2\boldsymbol{r}\ \psi^* \nabla^2 \psi
, \nonumber\\
E_{trap} &=&  \int d^2\boldsymbol{r}\ V(\boldsymbol{r})|\psi|^2 , \nonumber\\
E_{int} &=& \frac{1}{2}\frac{G_{2D}}{N} \int d^2\boldsymbol{r}\ |\psi|^4, \nonumber\\
E_{rot} &=& \int d^2\boldsymbol{r}\ \psi^* (\boldsymbol{\Omega}\cdot\boldsymbol{L}) \psi,
\end{eqnarray}
the kinetic, trapping potential, interaction, and rotation energies, respectively. They are related to the total energy $E$ through
\begin{eqnarray}
E=E_{kin}+E_{trap}+E_{int}-E_{rot}.
\end{eqnarray}

\begin{figure*}[!htb] 
\centering
\subfloat[\label{fig:shophase1a}]{\includegraphics[width=0.4\textwidth]{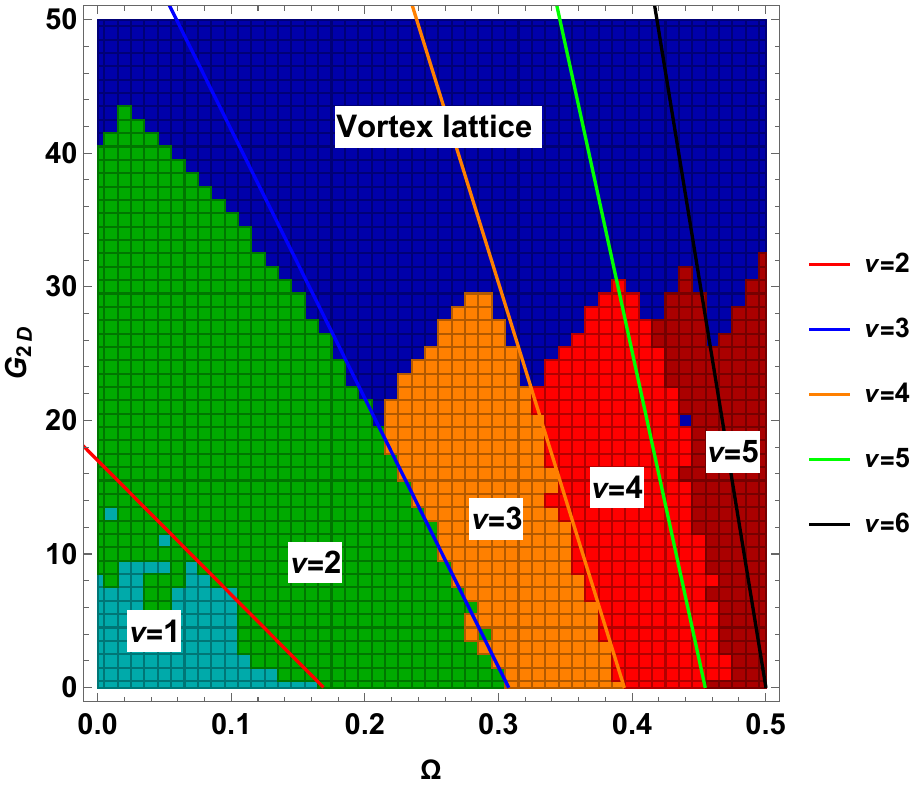}}
\subfloat[\label{fig:shophase1b}]{\includegraphics[width=0.125\textwidth]{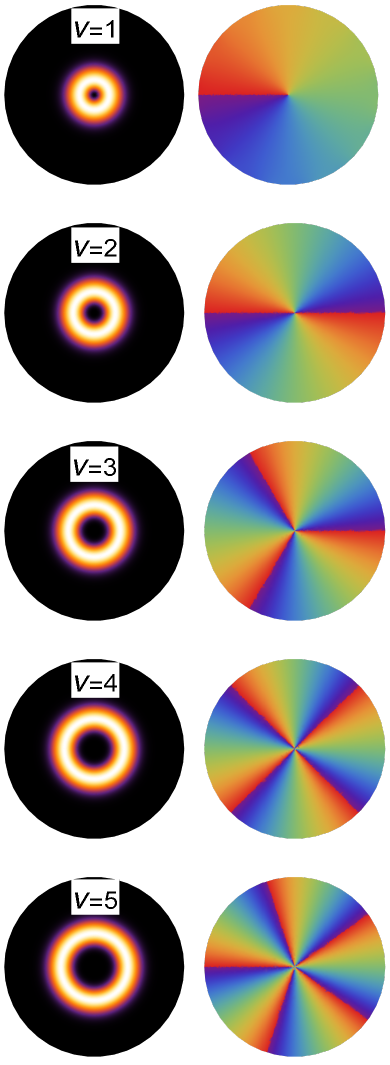}}
\subfloat[\label{fig:SHOPhase2a}]{\includegraphics[width=0.4\textwidth]{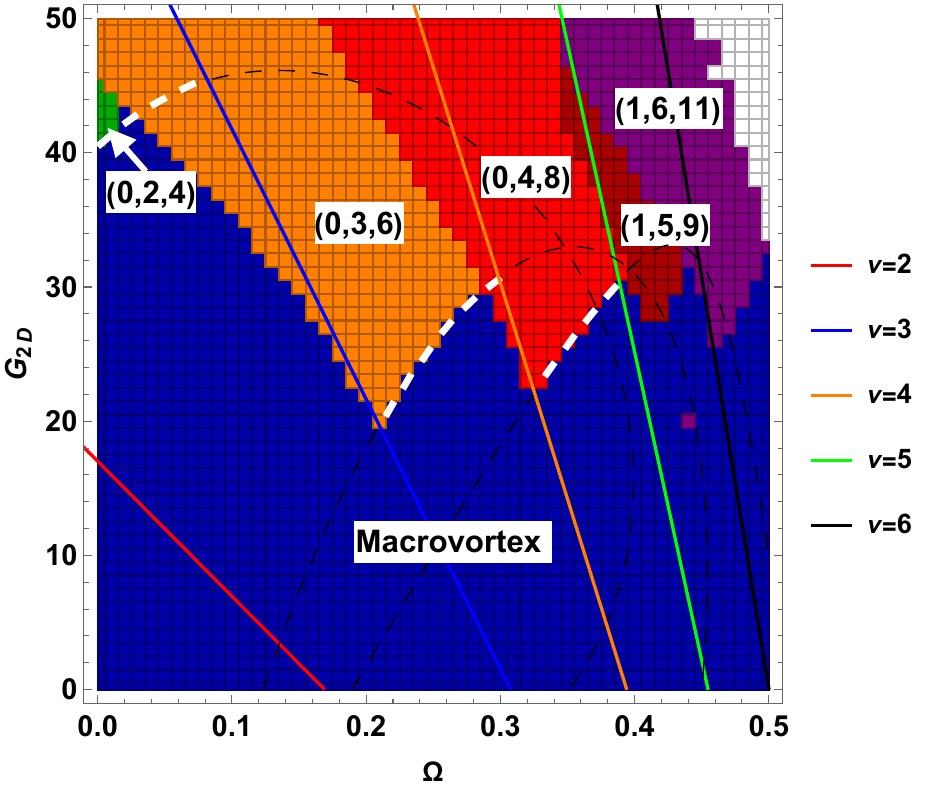}}
\subfloat[\label{fig:SHOPhase2b}]{\includegraphics[width=0.11\textwidth]{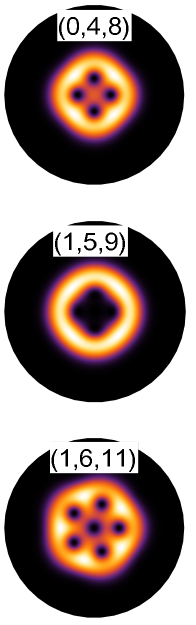}}
\caption{
Phase diagram of the SHO potential, Eq.~(\ref{eq:SHOPotential}), with $\rho_0=2.5$. The caption is the same as the one of Fig.~\ref{fig:MexPhase1}, but the first-order perturbation theory results for the macrovortex transitions are given by Eq.~(\ref{eq:sho_firstorder}). In contrast to Fig.~\ref{fig:MexPhase1}, the SHO potential appears to act as a pinning potential stabilizing the macrovortices, as these states occupy a larger area of the phase diagram when compared to the MH potential. In panel (c), the white color denotes a region where the numerical procedure is unstable, signalling its limit for large values of $\Omega$ and $G_{2D}$. The other colors define regions corresponding to different lattice regimes, indexed by the $(m_1,m_0,m_2)$ triples.
}
\label{fig:SHOPhase1}
\end{figure*}

The phase diagram analysis of these partial energies also gives insights into the transitions between different macrovortex states and vortex lattices. This can be seen in Fig.~\ref{fig:erot_mh}, which illustrates the quantized values of the rotational energy normalized by the frequency $\Omega$. In Fig.~\ref{fig:ekin_mh}, on the other hand, the variation in the kinetic energy appears to signal the same transitions predicted by the stability analysis. It decreases abruptly when we have the decay to a vortex lattice, which is more energetically favorable for larger values of the interaction parameter $G_{2D}$. We conclude that, in this trap, a multiply-charged vortex can be energetically stable for sufficiently weak interactions. However, as the interaction strength increases, it eventually becomes energetically favorable for the system to spread its angular momentum to additional states, and multiply-quantized vortex states break into singly-charged vortices. We found that the diagrams for $E_{trap}$ and $E_{int}$ did not yield any additional information about the transitions.

\subsection{Shifted harmonic oscillator}
\label{sec:sho}

For the SHO potential, we have Eq.~(\ref{rotFrame}) with
\begin{equation} \label{eq:SHOPotential}
V(\rho) = \frac{1}{2}\omega_{SHO}^2(\rho-\rho_0)^2. 
\end{equation}
It can be obtained from Eq.~(\ref{SHO_potentialapp}) with $\rho_0 = r_{\rm min}$ and $\omega_{SHO}^2 = 2/\sigma$. In this work, we fixed  $\sigma =2$.

Besides the change in the trapping potential, the variational numerical procedure is the same as the one described in Sec.~\ref{sec:mh_phase_diagram}. We employed the same range of $\Omega$ and $G_{2D}$ values as in the MH case to allow for direct comparison between the two traps. To avoid the instability of the SHO against rotation, in all our analyses, we always kept the rotation frequency below the harmonic confinement frequency $\Omega = \tilde{\Omega}/\tilde{\omega}_0 <1$.

Figure~\ref{fig:shophase1a} shows the phase diagram concerning the macrovortices. We observe that the macrovortex region fills a larger area of the phase diagram for the SHO when compared to the MH case. This reflects the fact that the SHO potential works as a stabilizer ``pinning'' potential of the macrovortex configuration, making it more stable against a continuous transition to the lattice phase if compared to the MH trap. Furthermore, we see that, for the same $\Omega$ used in the case of MH, we obtained higher values for the charge $\nu$ of the macrovortex. Thus, we conclude that with the shifted harmonic oscillator trap, we could transfer angular momentum more efficiently to the superfluid, reaching higher vorticities with lower $\Omega$ values and having a more stable macrovortex configuration to be studied experimentally. In Fig.~\ref{fig:shophase1b}, we reproduce the density and phase profiles of the observed macrovortices.

To accurately describe the regions with large values of $\Omega$ and $G_{2D}$, we had to carefully check the number of terms used in the variational calculation, Eq.~(\ref{ansatzMH}). We employed $N_{\rm max}\sim 3 \nu$ in the case of macrovortices and $N_{\rm max}\sim 3 m_0$ for vortex lattices. For a macrovortex, $N_{\rm max}=\nu$ would suffice, but for a vortex lattice described by the triple $(m_1,m_0,m_2)$, we need $N_{\rm max}$ to be at least the maximum between these three values. For a fixed value of $m_0$, the $m_1+m_2=2m_0$ constrain makes the triples of the form $(0,m_0, 2 m_0)$ the more demanding ones (for $m_1\neq 0$ then $m_2<2m_0$ and fewer coefficients are required). Hence we employed $N_{\rm max}\sim 3 \nu$ or $\sim 3 m_0$ to have a margin of safety and also because, given a $(\Omega,G_{2D})$ pair, we do not know \textit{a priori} what the vortex configuration is.

Using first-order perturbation theory, we determined the energy for a pure macrovortex state characterized by a single number $\nu$,
\begin{equation}  \label{OmegaCrit2}
E_\nu = 1+\nu+\frac{\rho_0^2}{2}-\frac{\rho_0 \sqrt{\pi}}{2}\frac{(2\nu+1)!}{2^{2\nu}(\nu!)^2} +\frac{G_{2D}}{4\pi}\frac{(2\nu)!}{2^{2\nu}(\nu!)^2} -\nu\Omega.
\end{equation} Equating the energies $E_\nu$ and $E_{\nu+1}$, we obtain the critical frequency for the macrovortex phase transition,
\begin{equation}
\label{eq:sho_firstorder}
\Omega\left(G_{2D}\right) = 1-\frac{\rho_0\sqrt{\pi}}{2}\frac{(2\nu)!}{2^{2\nu}(\nu!)^2} - \frac{G_{2D}}{2\pi}\frac{(2\nu-2)!}{2^{2\nu}(\nu-1)!\nu!}.
\end{equation}
We included the lines given by Eq.~(\ref{eq:sho_firstorder}) in Fig.~\ref{fig:shophase1a}. We observe a qualitative agreement for $\nu< 4$, but for higher values of the macrovortex charge, the regions are not well-described by first-order perturbation theory.

The absence of the $\nu=0$ mode is a consequence of Eq.~(\ref{OmegaCrit2}), which shows that for any value of $\Omega$, $E_0$ is always larger than $E_1$. This equation was derived using an ansatz where a hole cannot be present in the center of the condensate for zero angular momentum, so it is natural to expect the absence of a zero angular momentum region. Certainly, the mode $\nu =0$ exists in the SHO case for $\Omega = 0$. The variational method, however, does not give this particular phase for the SHO due to the chosen ansatz. Since our interest is in rotating condensates, this region should not affect our analysis.

\begin{figure*}[!htb] 
    \centering
    \begin{minipage}{0.33\textwidth}
        \centering
        \subfloat[]{\includegraphics[width=0.9\textwidth]{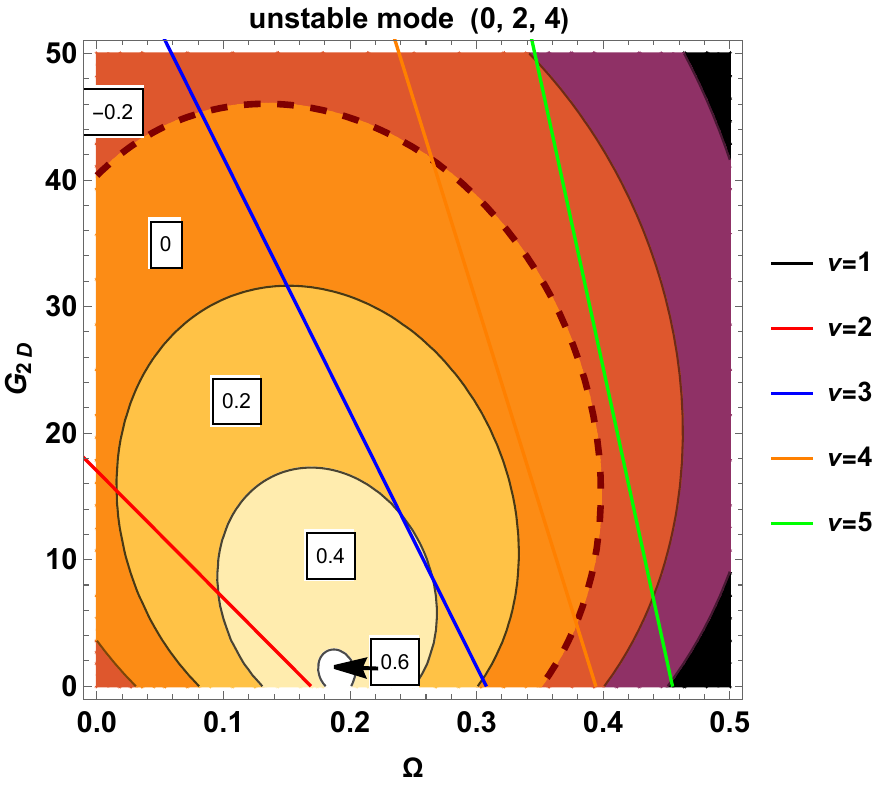}}
    \end{minipage}\hfill
    \begin{minipage}{0.33\textwidth}
        \centering
        \subfloat[]{\includegraphics[width=0.9\textwidth]{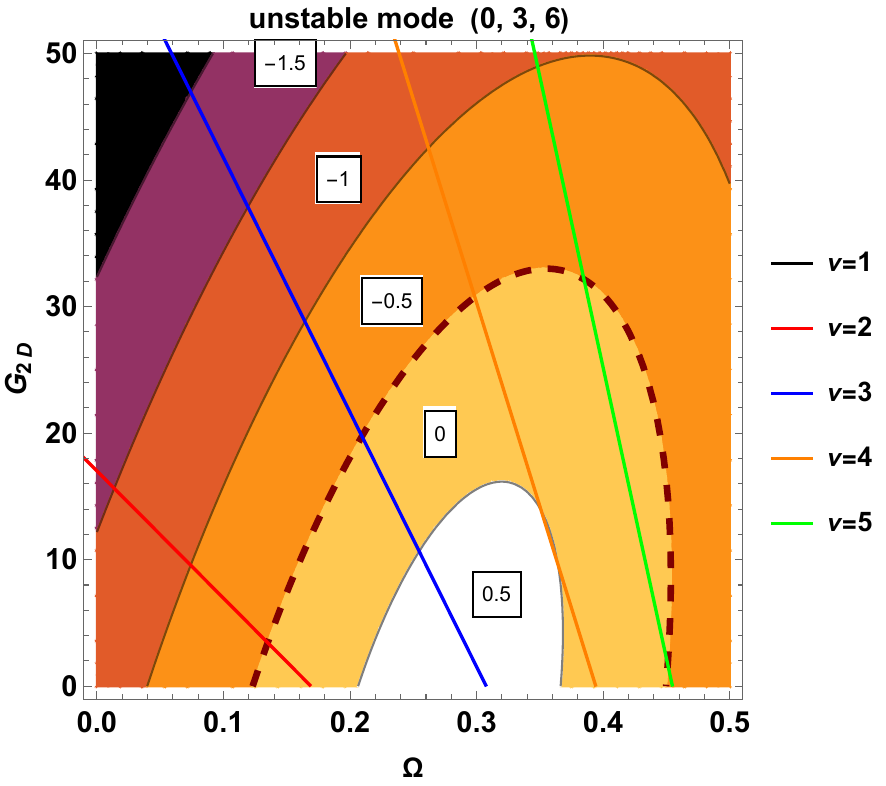}}
    \end{minipage}
    \begin{minipage}{0.33\textwidth}
        \centering
        \subfloat[]{\includegraphics[width=0.9\textwidth]{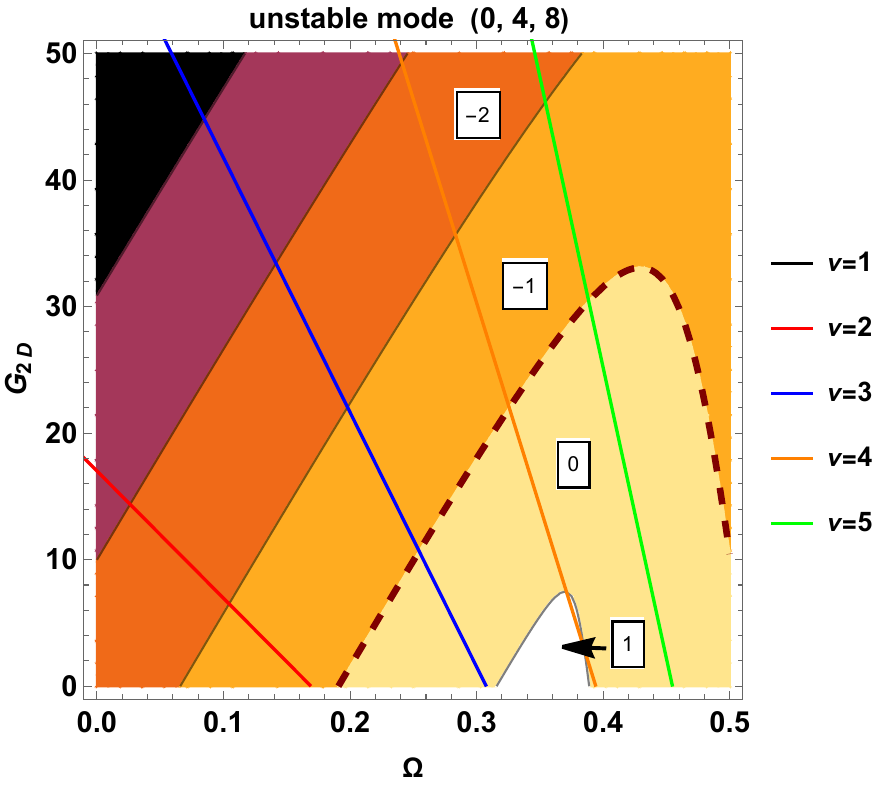}}
    \end{minipage}\hfill
      \caption{The figures indicate the value of the smallest eigenvalue of the stability matrix for the pure macrovortex states $m_0 =2,3,4$. The boundary of the positive region in each diagram corresponds to a phase transition from macrovortex to vortex lattice regime.}
      \label{fig:SHO2ndOrder}
\end{figure*}

In Fig.~\ref{fig:SHOPhase2a} we focus on the vortex lattices. Because of angular momentum conservation, as in the case of MH, we have the constraint $m_1+m_2=2m_0$, with $m_0$ having majority occupation and $|c_{m_0}|^2+|c_{m_1 }|^2+|c_{m_2}|^2 \approx 1$ (within a $10^{-3}$ tolerance). The blue color characterizes the macrovortices region and each other color indicates a vortex lattice, indexed by the $(m_1,m_0,m_2)$ triples. The correspondent lattices density profiles are given in Fig.~\ref{fig:SHOPhase2b}. The analysis based on the stability matrices calculated for each pure state mode, as shown in Fig.~\ref{fig:SHO2ndOrder}, accurately predicted the boundary between the macrovortex and lattice regions.

In the MH case, we only observed lattices of the $(0,m_0,2m_0)$ type. For the SHO trapping potential, we notice the presence of the vortex lattices $(1,5,9)$ and $(1, 6, 11)$, which have a central vortex of charge one.

For large values of $\Omega$ and $G_{2D}$, the white region in Fig.~\ref{fig:SHOPhase2a}, we observed a non-negligible value for more than three $c_m$ coefficients. Increasing the number of $N_{\rm max}$ terms used in the variational calculation, Eq.~(\ref{ansatzMH}), led to numerical instabilities. This indicates the limitation of the procedure we adopted, as it is, for large values of $\Omega$ and $G_{2D}$. In future studies, we intend to investigate if the parameter space can be constrained to improve the variational procedure, if a different ansatz is necessary for this region, or if more than three $c_m$ coefficients truly characterize the phase.

\begin{figure}[!htb] 
      \centering
    \begin{minipage}{0.35\textwidth}
        \raggedleft
        \subfloat[]{\includegraphics[width=\textwidth]{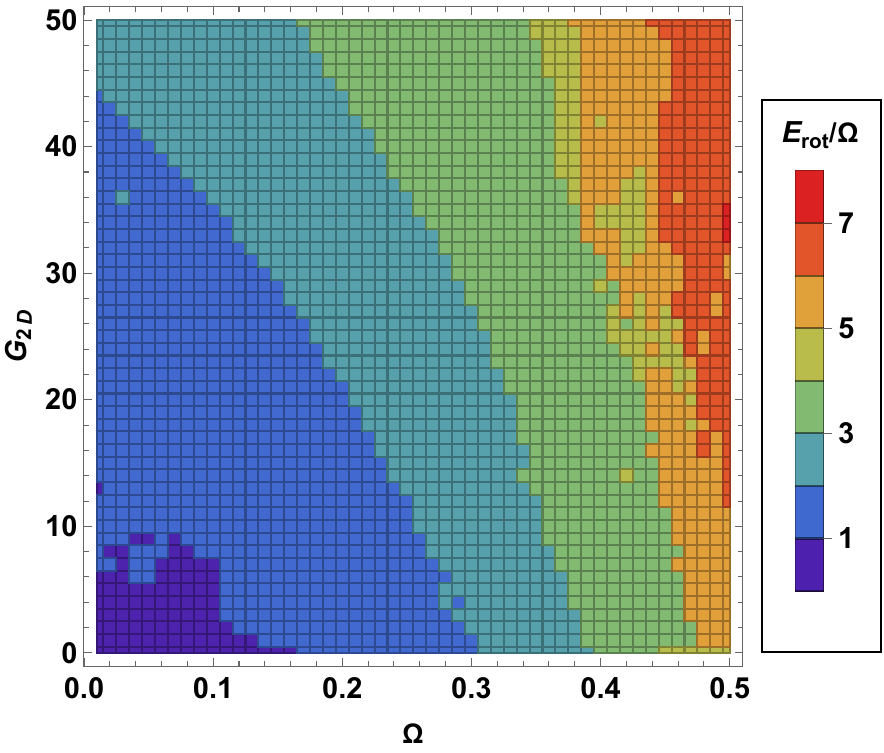}}
    \end{minipage}\hfill
    \begin{minipage}{0.35\textwidth}
        \centering
        \subfloat[]{\includegraphics[width=\textwidth]{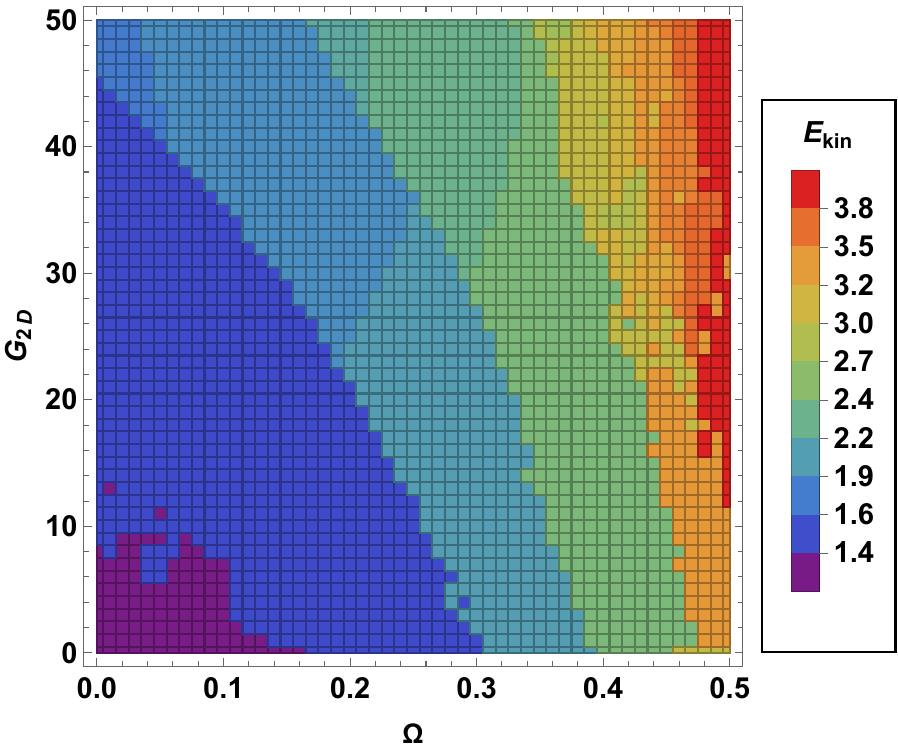}}
    \end{minipage}
      \caption{
Partial energy diagrams for the SHO potential, Eq.~(\ref{eq:SHOPotential}), with $\rho_0=2.5$. (a) The rotational energy indicates the phase transitions between macrovortices with different charges. (b) In the kinetic energy, it is possible to see the transitions from macrovortices to vortex lattices.
      \label{fig:SHOEnergies}}
\end{figure}

Finally, in Fig.~\ref{fig:SHOEnergies}, we present the analysis of the partial energy diagrams. As in the case of the MH potential, the rotational energy normalized by $\Omega$ shows the quantized values of $\nu$ related to the phases of macrovortices. On the other hand, the kinetic energy diagram shows a sudden decrease in energy when crossing the boundary from a macrovortex to a vortex lattice configuration. 

The energy analysis is explored further in the next section, where we show how the frequencies of the collective modes change when the system evolves from a macrovortex to a singly-charged vortex lattice configuration.

\section{Collective modes and Sum Rules}\label{sec: COllective modes}

In this section, through the sum rules method, we derived expressions for the frequencies of the collective modes in terms of the moments of the strength function. These are analogous to the results of Ref.~\cite{Cozzini2006b}. However, here we calculate the moments for different confinement traps (MH and SHO), which yield new dependencies of the frequencies with respect to the partial energies of the system.

To investigate the collective modes of a system described by $\hat{H}$, we consider a complete set of exact eigenstates $\{| n \rangle \}$ and eigenvalues $\{E_n\}$ of this Hamiltonian, with $E_0\leqslant E_1\leqslant E_2 ...$, and also an excitation operator of the system $\hat{F}$. An upper bound $\hbar\omega^{\rm upper}$ to the minimum excitation energy $\hbar\omega^{\rm min}=E_1-E_0$ of the states excited by $\hat{F}$ is given by
\begin{equation} \label{bound}
\hbar \omega^{\rm upper} \geqslant \sqrt{m_3/m_1},
\end{equation}
where one has the moments $m_p$ of the strength function $S(\omega)$,
\begin{eqnarray} \nonumber \label{momentsH}
m_p&&= \int_{0}^{\infty} S(\omega) \omega^p d\omega \\
&&= \langle 0|\hat{F}^{\dagger}(\hat{H}-E_0)^p\hat{F}|0 \rangle.
\end{eqnarray}
From this definition of the moments and the strength function [see Eq.~(\ref{eq:strength_function_app})], one can show that
\begin{eqnarray}
\frac{(\omega^{\rm upper})^2-(\omega^{\rm min})^2}{(\omega_1-\omega_0)^2}+1&=&\frac{\sum\limits_{n>0}|\langle n|\hat{F}|0 \rangle|^2 \frac{(\omega_n-\omega_0)^3}{(\omega_1-\omega_0)^3}}{\sum\limits_{n>0}|\langle n|\hat{F}|0 \rangle|^2 \frac{(\omega_n-\omega_0)}{(\omega_1-\omega_0)}}. \nonumber \\
&&
\end{eqnarray} 
Since we have
\begin{equation}
\frac{(\omega_n-\omega_0)}{(\omega_1-\omega_0)}\geqslant 1,
\end{equation}  we conclude that $\omega^{\rm upper} \geqslant  \omega^{\rm min}$. The advantage of the sum-rule approach is that just the ground state is required to calculate the moments [see Eq.~(\ref{momentsH})] and then to find the strength function. If, in particular, the excitation operator $\hat{F}$ excites a single state and we have knowledge of the exact ground state, $\hbar\omega^{\rm upper}$ gives the exact excitation energy.

Before we introduce the collective modes, which are characterized by the symmetries of the chosen excitation operator $\hat{F}$, we need to generalize the definition of the moments for a non-Hermitian operator ($\hat{F} \neq \hat{F}^{\dagger}$). Calling  $\hat{F}=F_+$ and $\hat{F}^{\dagger}=F_-$,  we define
\begin{equation} \label{strength}
m_p^{\pm}=\int_{0}^{\infty} \; d\omega \left[S_+(\omega)\pm S_-(\omega)\right] \omega^p,
\end{equation} with
\begin{equation}
S_{\pm}(\omega)= \sum_{n>0}  |\langle n|F_{\pm}|0 \rangle|^2 \delta(\omega-\omega_{\pm}),
\end{equation} where $\omega_{\pm}$ corresponds to the energies of the modes that we want to determine. In terms of the previous definition given by Eq.~(\ref{momentsH}), we have that $m_p^{\pm}=m_p(\hat{F})\pm m_p(\hat{F}^{\dagger})$. We will apply the general formula in Eq.~(\ref{strength}) to deal with the quadrupole mode operator.

\subsection{Monopole mode} \label{subsec:monopole}

Previously, we considered the upper bound of the lowest energy excited by an operator $\hat{F}$. The equality in Eq.~(\ref{bound}) holds whenever $\hat{F}$ excites one single mode. This section studies the breathing oscillation expected to be excited by the monopole operator. We extracted the frequency of the lowest excited mode from
\begin{equation} \label{Wupper}
\hbar \omega = \sqrt{m_3/m_1}.
\end{equation} The explicit calculation of the moments $m_p$ for $p>0$ can be carried out in terms of commutators between the excitation operator and the Hamiltonian of the system evaluated on the ground state \cite{LIPPARINI1989},
\begin{equation}
m_1= \frac{1}{2}\langle 0 | [\hat F,[\hat H,\hat F]]|0 \rangle,
\end{equation}
and
\begin{equation}
m_3= \frac{1}{2}\langle 0 | [[\hat F,\hat H],[\hat H,[\hat H,\hat F]]]|0 \rangle.
\end{equation} In the following, we considered the perturbation given by the 2D version of the monopole operator [see Eq.~(\ref{FM})] and the rotating frame many-body Hamiltonian, Eq.~(\ref{rotFrame}).

For the Mexican hat trapping potential with anharmonicity $\lambda$ we obtain the frequency
\begin{equation}\label{FmMH}
\omega_{{M}}^{MH} =  \sqrt{2\frac{E_{kin}}{E_{ho}}-2+2\frac{E_{int}}{E_{ho}}+8 \lambda  \frac{\langle \rho^4 \rangle}{\langle \rho^2 \rangle}}.
\end{equation} 
Using the Virial theorem (see appendix \ref{sec:ap1}), it can be rewritten as
\begin{equation}\label{FmMH2}
\omega_{{M}}^{MH} = \sqrt{2+6\frac{E_{kin}}{E_{ho}}+6\frac{E_{int}}{E_{ho}}}.
\end{equation}
Repeating the same calculation for the shifted harmonic oscillator trap with radius $\rho_0$ yields
\begin{equation} \label{FmSHO}
\omega_{{M}}^{\rm SHO} =  \sqrt{2\frac{E_{kin}}{E_{ho}}+2+2\frac{E_{int}}{E_{ho}}- \rho_0\frac{ \langle \rho \rangle}{\langle \rho^2 \rangle}},
\end{equation}
which is equivalent to
\begin{equation}\label{FmSHO2}
\omega_{{M}}^{\rm SHO} = \sqrt{1+3\frac{E_{kin}}{E_{ho}}+3\frac{E_{int}}{E_{ho}}}.
\end{equation}

Equations~(\ref{FmMH})-(\ref{FmSHO2}) contain the kinetic and interaction energies given by the same expressions as in Eq.~(\ref{eq:partial_energies}). The difference is that in Secs.~\ref{sec:second_order_mh} and \ref{sec:sho}, we used the variational wave function to compute the partial energies, but here we employed the ground state solution of the GPE. Equations~(\ref{FmMH})-(\ref{FmSHO2}) also feature $E_{ho}$, given by
\begin{equation} \nonumber
E_{ho}=\frac{\langle \rho^2 \rangle}{2}.
\end{equation}

\subsection{Quadrupole mode}  \label{subsec:quadrupole}

To calculate the quadrupole modes in a condensate with vortices, we have to include a larger number of modes and moments in the analysis.
Importantly, a splitting in the quadrupole mode frequency is expected due to vortices in the condensate ~\cite{Zambelli1998}.
This can be understood by noting that the average velocity flow associated with the collective oscillation can be either parallel or opposite to the vortex flow, depending on the sign of the angular momentum carried by the excitation, thus producing a shift in the collective mode frequency. Specifically, to deal with the frequency degenerescence in the sum rule, we used the two-modes approach, 
\begin{equation}
S_{\pm}(\omega)= \sigma^{\pm} \delta(\omega-\omega_{\pm}),
\end{equation} with $\sigma^+=\sigma^-=\sigma_0$ ($m_0^-=\sigma^+ - \sigma^-=0$). Employing this method, the frequencies are given by (for the details, see Appendix~\ref{sec:app_sum_rules}),

\begin{equation} \label{twomodes}
\omega_{\pm}=\frac{1}{2}\left[\sqrt{\left(\frac{m_2^-}{m_1^+}\right)^2+4\left(\frac{m_4^-}{m_2^-}-\frac{m_3^+}{m_1^+}\right)} \mp \frac{m_2^-}{m_1^+}\right].
\end{equation}

As we will see afterwards, when we have the annular structure of the condensate, both the low- ($\omega_-$) and high-lying ($\omega_+$) energy branches acquire additional modes. These new modes, which have opposite azimuthal quantum numbers $m$, are very close in energy to the original $m = \pm 2$ modes. A full treatment of the 4-mode system would be very complicated. Instead, we assume that two doubly-degenerate energy levels are present in our analysis. Then, in the sum-rule calculation, we consider Eq.~(\ref{strength}) with
\begin{eqnarray}
S_+(\omega) &=& \sigma_H \delta(\omega - \omega_H) + \sigma_L  \delta(\omega - \omega_L),\nonumber\\
S_-(\omega) &=& \sigma_H^{\dagger} \delta(\omega - \omega_H) + \sigma_L^{\dagger}   \delta(\omega - \omega_L).
\end{eqnarray}
That is, we assume the degenerescence between the energies of the higher ($\sigma_H$ and $\sigma_H^{\dagger}$) and lower ($\sigma_L$ and $\sigma_L^{\dagger}$) branches of the $m = \pm 2$ modes~\cite{Cozzini2006b}.

This gives us the solutions (see Appendix~\ref{sec:app_sum_rules}),
\begin{equation} \label{fourmodes}
\omega_{H,L}^2=\frac{1}{2}\left[\frac{m_4^-}{m_2^-} \pm \sqrt{\left(\frac{m_4^-}{m_2^-}\right)^2-4 \frac{m_{1}^+}{m_{-1} }\left(\frac{m_4^-}{m_2^-}-\frac{m_3^+}{m_1^+}\right)}\right].
\end{equation}

To calculate the two- and four-modes frequencies in Eqs.~(\ref{twomodes}) and (\ref{fourmodes}), we considered the non-Hermitian quadrupole perturbation operator as in Eq.~(\ref{FQ}), \begin{equation}
F_{\pm}= \sum_{j=1}^N (x_j\pm i y_j)^2,
\end{equation} where $F_+=F_-^{\dagger}$. The details of the calculations of the frequencies of the modes are given in Appendix \ref{ap:moments_calculations}.

\subsection{Comparison with numerical simulations}
\label{sec:numerial_simulations}

Finally, to complete our analysis of the collective modes, we considered numerical solutions of the Gross-Pitaevski equation,
\begin{equation} \label{eq:GPE}
i\frac{\partial \psi}{\partial t}=-\frac{1}{2}\nabla^2\psi+V\psi+G_{2D}|\psi|^2\psi-\boldsymbol{\Omega}\cdot\boldsymbol{L}\psi,
\end{equation}
where $\nabla^2$ is the two-dimensional Laplacian operator. The propagation in imaginary time yields the ground state, while real time simulations can be employed to study the dynamics of the condensate. We modified the code developed in Ref.~\cite{Kumar2019}, which uses a Crank-Nicolson scheme to find the numerical wave function.

\begin{figure}[!htb] 
    \centering
    \begin{minipage}{0.5\textwidth}
        \centering
        \subfloat[]{\includegraphics[width=1.0\textwidth]{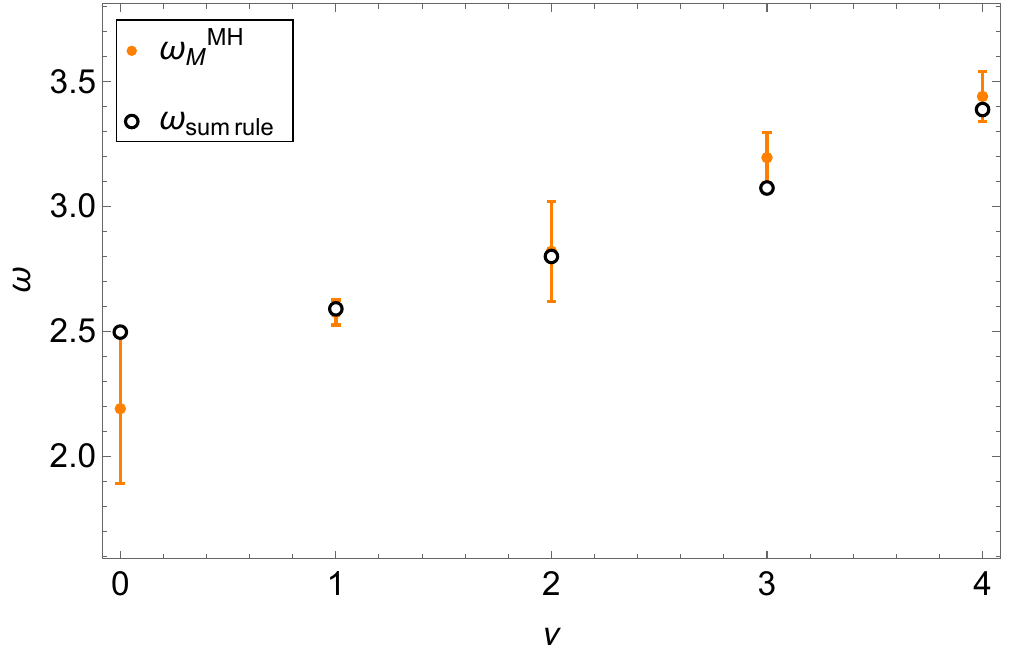}}
    \end{minipage}\hfill
    \begin{minipage}{0.5\textwidth}
        \centering
        \subfloat[]{\includegraphics[width=1.0\textwidth]{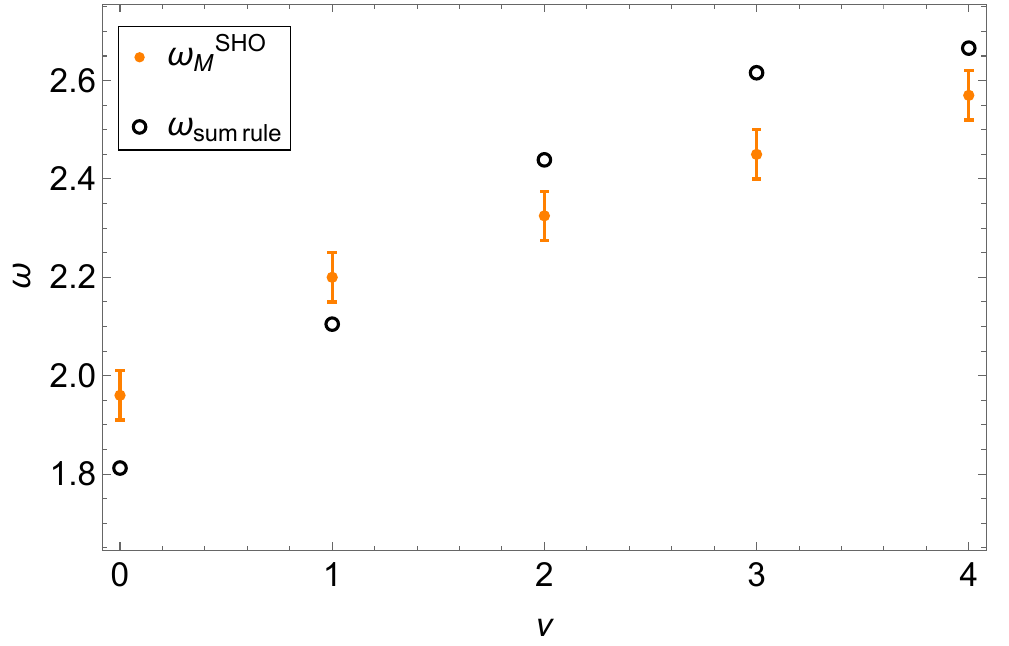}} 
    \end{minipage}
      \caption{Comparison between the monopole frequency obtained with the numerical solutions of the GPE and the results of the sum rules predictions as given by Eq.~(\ref{FmSHO}). (a) MH with $\lambda=0.05$ and (b) SHO  with $\rho_0=2.5$, both with the same interaction strength $G_{2D}=10$. The frequency increases with larger vorticities, indicating that vortices reduce the compressibility of the gas. \label{fig:monopoleSHO}}
    \end{figure}

\begin{figure*}[!htb]
    \centering
    \begin{minipage}{0.5\textwidth}
        \centering
        \subfloat[]{\includegraphics[width=1.0\textwidth]{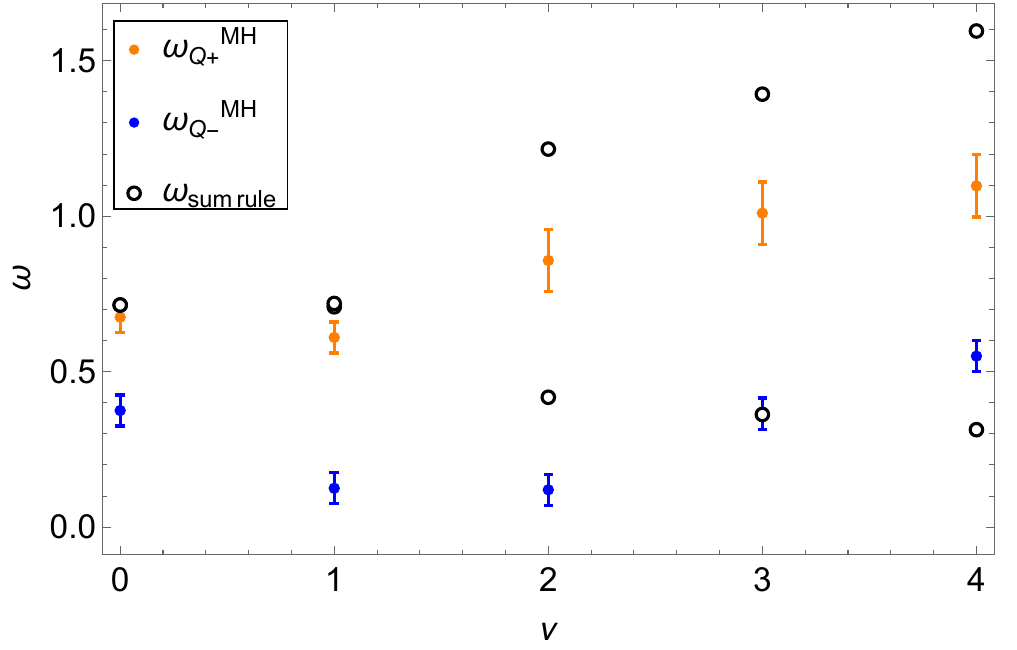}}
    \end{minipage}\hfill
    \begin{minipage}{0.5\textwidth}
        \centering
        \subfloat[]{\includegraphics[width=1.0\textwidth]{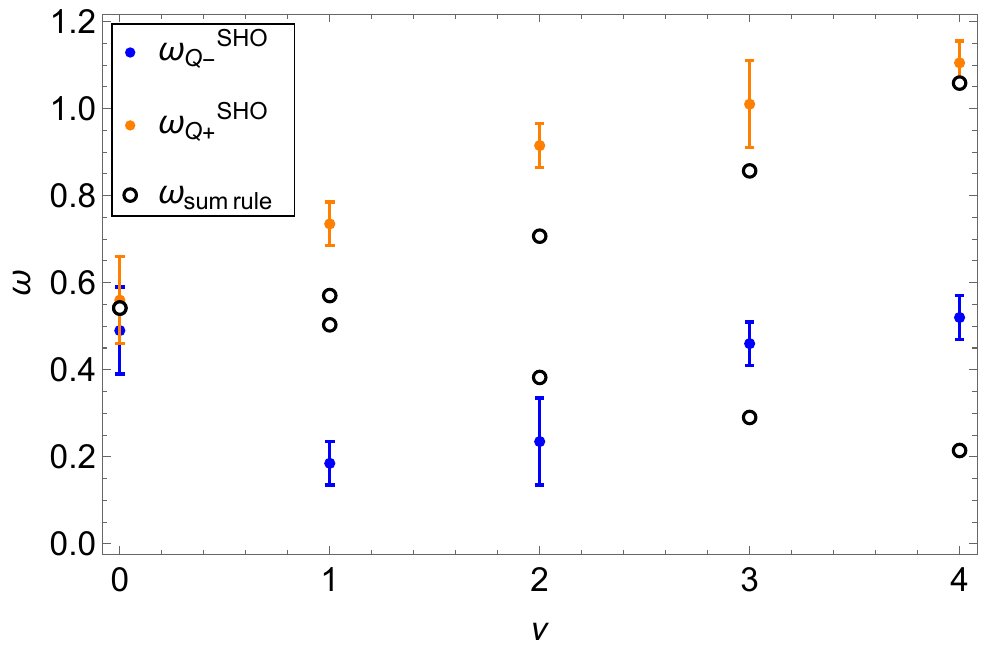}}
    \end{minipage}
    \centering
    \begin{minipage}{0.5\textwidth}
        \centering
        \subfloat[]{\includegraphics[width=1.0\textwidth]{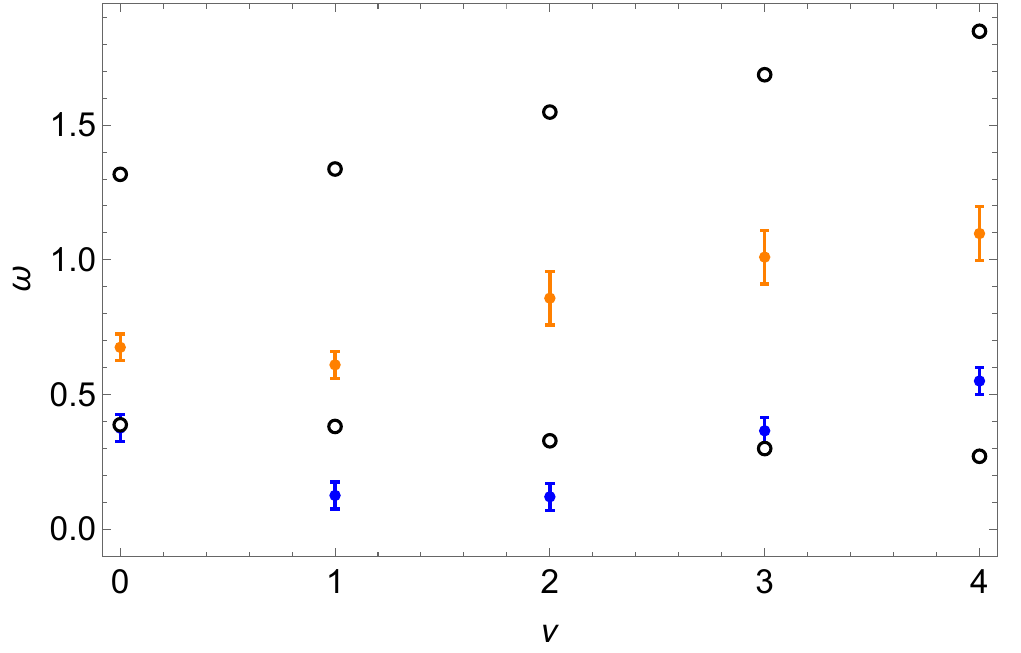}}
    \end{minipage}\hfill
    \begin{minipage}{0.5\textwidth}
        \centering
        \subfloat[]{\includegraphics[width=1.0\textwidth]{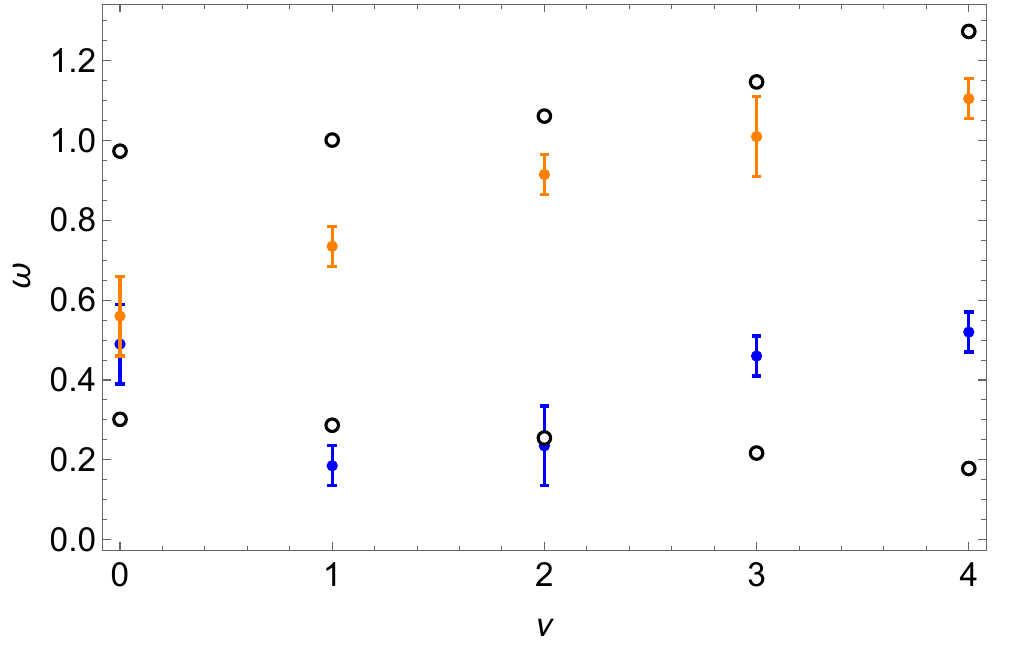}} 
    \end{minipage}
         \caption{
Comparison between the quadrupole frequency obtained with the numerical solution of the GPE, the results of the (a-b) two-modes quadrupole sum rule, Eq.~(\ref{twomodes}), and the (c-d) four-modes quadrupole sum rule,  Eq.~(\ref{fourmodes}). Panels (a) and (c) contain the results regarding the MH potential with $\lambda=0.05$, and (b) and (d) stand for the SHO potential with $\rho_0=2.5$. In both cases, we employed the same interaction strength $G_{2D}=10$. We can see that the splitting between the modes increases with increasing $\nu$, being proportional to the angular momentum. The four-modes approach better reproduces the numerical results. 
\label{fig:quadrupole4mode}}
    \end{figure*}

Our variational results obtained in Sec.~\ref{sec:diagram}, for a given $\Omega$ and $G_{2D}$,  were used as the initial conditions for imaginary time propagation. This choice led to steady states with greater accuracy and shorter convergence time. Notably, the wave function phase is the most relevant input to the GPE calculations since its amplitude is adjusted during the numerical simulations. The resulting numerical ground state of the system is then used both for computing the average values described in Secs.~\ref{subsec:monopole} and \ref{subsec:quadrupole} and for investigating the collective modes in real-time propagation.

\begin{figure*}[!htb]
    \centering
    \begin{minipage}{0.33\textwidth}
        \centering
        \subfloat[]{\includegraphics[width=1.0\textwidth]{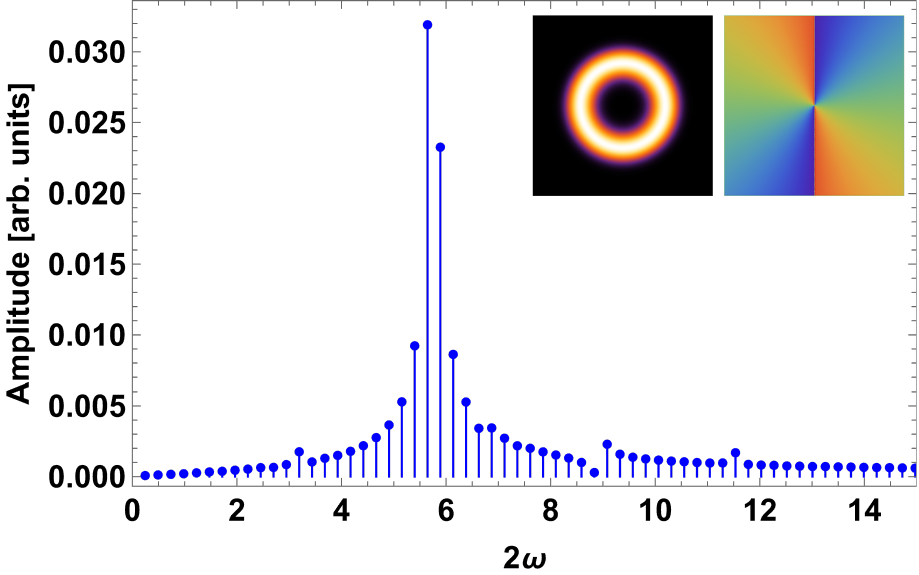}}
    \end{minipage}\hfill
    \begin{minipage}{0.33\textwidth}
        \centering
        \subfloat[]{\includegraphics[width=1.0\textwidth]{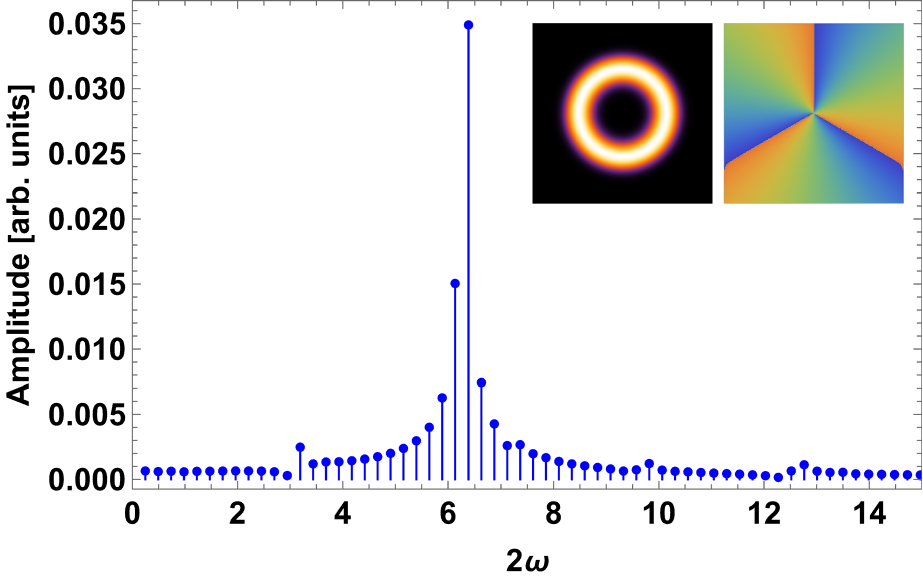}}
    \end{minipage}
     \begin{minipage}{0.33\textwidth}
        \centering
        \subfloat[]{\includegraphics[width=1.0\textwidth]{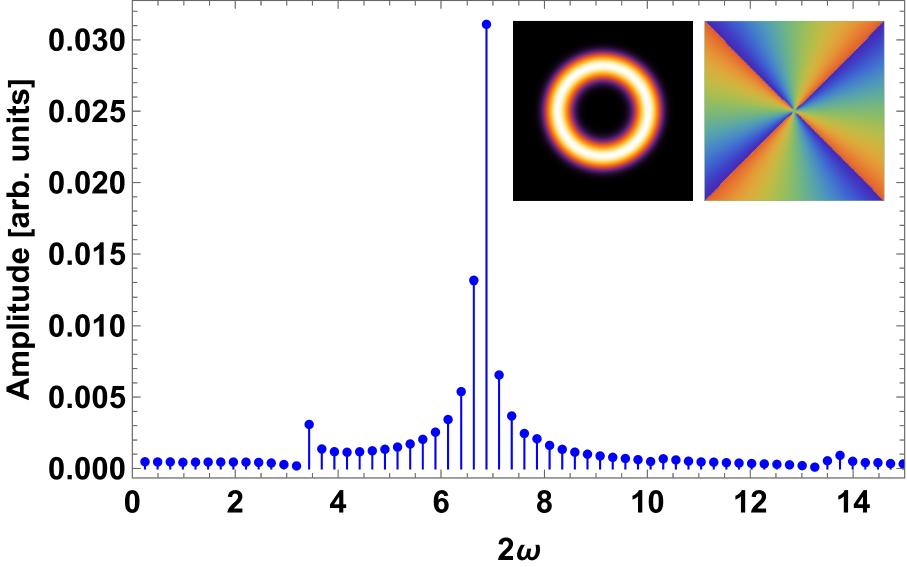}}
    \end{minipage}
        \begin{minipage}{0.33\textwidth}
        \centering
        \subfloat[]{\includegraphics[width=1.0\textwidth]{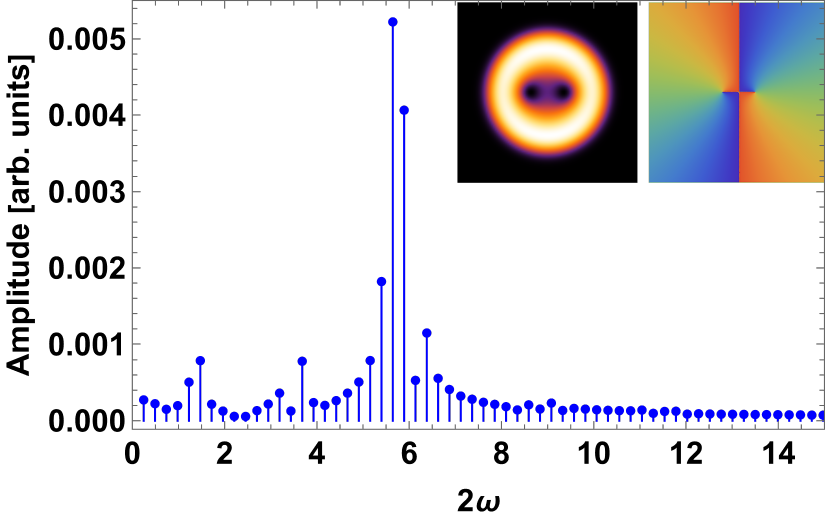}}
    \end{minipage}\hfill
    \begin{minipage}{0.33\textwidth}
        \centering
        \subfloat[]{\includegraphics[width=1.0\textwidth]{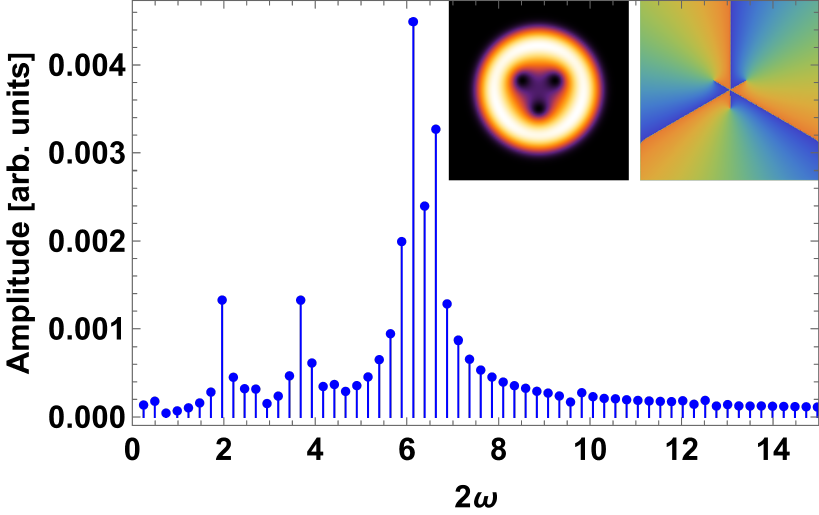}}
    \end{minipage}
     \begin{minipage}{0.33\textwidth}
        \centering
        \subfloat[]{\includegraphics[width=1.0\textwidth]{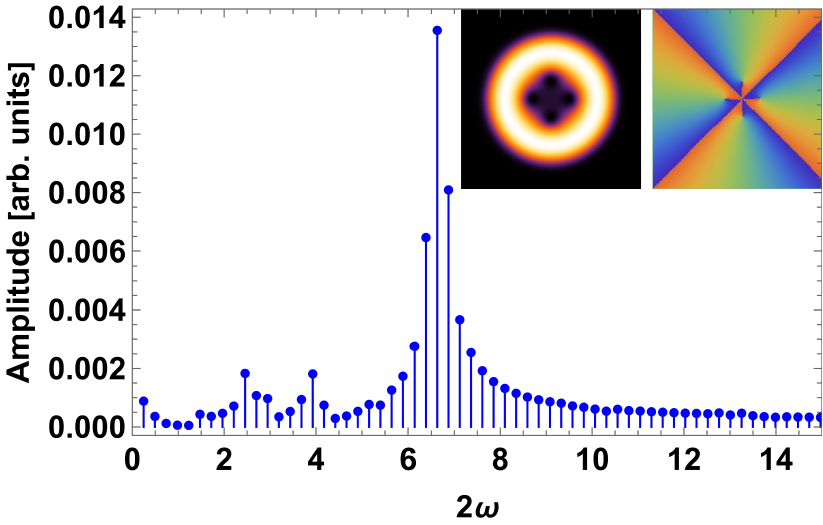}}
    \end{minipage}
         \caption{
Monopole mode of the Mexican hat trap with $\lambda =0.05$. The interaction parameter is $G_{2D}=10$ for the (a-c) macrovortices and $G_{2D}=100$ for the (d-f) vortex lattices. The rotation frequencies are $\Omega=0.3$, $0.4$, and $0.5$ for panels (a) and (d), (b) and (e), (c) and (f), respectively. In the insets, we reproduce the density and phase profiles of the ground state. \label{fig:FFT_MH}}
    \end{figure*}

A straightforward way of exciting collective modes is considering slightly different Hamiltonians in the imaginary and real time simulations. This means that the ground state obtained in the imaginary time simulation is not an eigenstate of the Hamiltonian used for the real time evolution. The subsequent dynamics are used to probe the collective modes. To observe the monopole mode, we increased the value of $G_{2D}$ by 20\%. We introduced a small anisotropy in the trapping potential for the quadrupole mode, increasing the frequency in the $x$ direction by 5\%. For each $\Omega$, $G_{2D}$, collective mode and trapping potential, we computed the wave function in $1024$ different instants, forming a time series. We extracted the frequency of the monopole mode $\omega_{\mathrm{M}}$ by performing a Fourier analysis of $\langle x^2 \rangle$. Analogously, we obtained the quadrupole mode frequency $\omega_{\mathrm{Q}}$ by considering $\langle x^2-y^2 \rangle/\langle x^2+y^2 \rangle$.

In the following, we considered the MH with anharmonicity $\lambda=0.05$ and the SHO trap with the shift $\rho_0=2.5$. The resolution of the FFT technique gives the error bars in the numerical results. The predictions using sum rules correspond to the expressions of Eqs.~(\ref{FmMH2}) and (\ref{FmSHO2}) for the monopole, and Eqs.~(\ref{twomodes}) and Eq.(\ref{fourmodes}) for the quadrupole modes. These are functions of the partial energies of the system, which we calculated with the numerical ground state solution obtained with the imaginary time propagation of the GPE.

In Fig.~\ref{fig:monopoleSHO}, we show the monopole frequency of the macrovortex configuration as a function of the vorticity $\nu$, for the MH and SHO traps, with a fixed interaction strength $G_{2D}=10$. The sum rule prediction reasonably agrees with the numerical results for the MH and the SHO potential. 

Regarding the quadrupole results, Fig.~\ref{fig:quadrupole4mode}, we see the increase in the modes separation with the macrovortex vorticity $\nu$. The four-modes sum rule prediction, Eq.~(\ref{fourmodes}), is better suited to reproduce the numerical data when compared to the two-modes approach. However, for the MH trap, we observe the sum rule overestimation of the numerical results of the high-frequency branch. In the MH case, the higher modes coupling due to the anharmonic term is not negligible, indicating that we should include more modes in the sum rules derivation. It is worth remembering that the four-modes approach also accounted for the additional modes that appear due to the existence of two walls in the ring configuration, with very close frequencies, both showing the separation into its components $m = \pm 2$ as we increase the vorticity of the condensed cloud. However, as these modes have very close energy values, it is difficult to distinguish them, mostly for large $\Omega$, so we only include two energy branches, each representing the mixture of the $+2$ and $-2$ components of each wall mode.

It is important to note that Eq.~(\ref{fourmodes}) depends on the compressibility moment $m_{-1}$. Unlike the other modes, this particular moment is not described in terms of partial energies. Its calculation is based on linear response theory, with the proper perturbation for each condensate mode, as detailed in Appendix~\ref{compressibility}.

\begin{figure*}[!htb] 
    \centering
    \begin{minipage}{0.33\textwidth}
        \centering
        \subfloat[]{\includegraphics[width=1.0\textwidth]{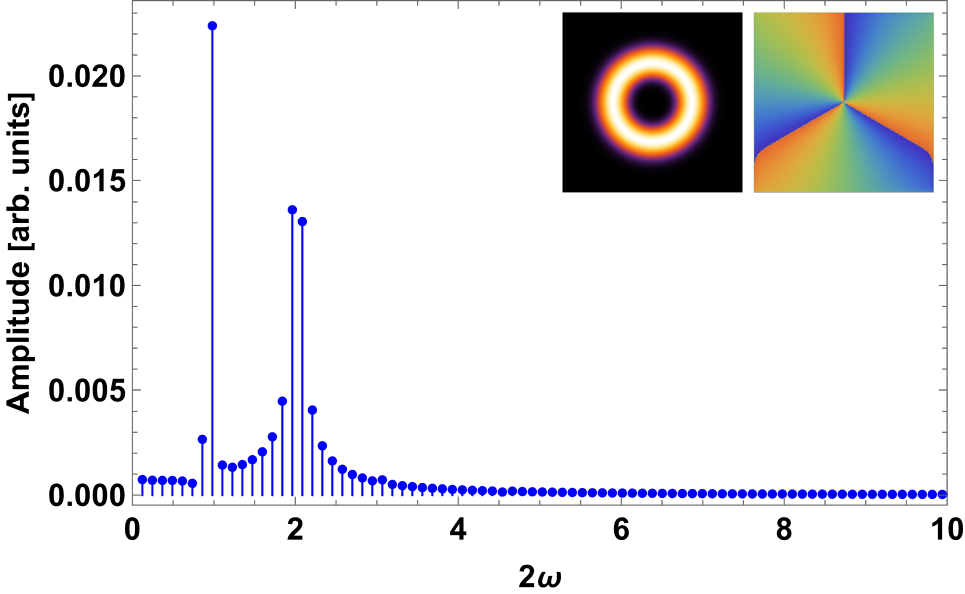}}
    \end{minipage}\hfill
    \begin{minipage}{0.33\textwidth}
        \centering
        \subfloat[]{\includegraphics[width=1.0\textwidth]{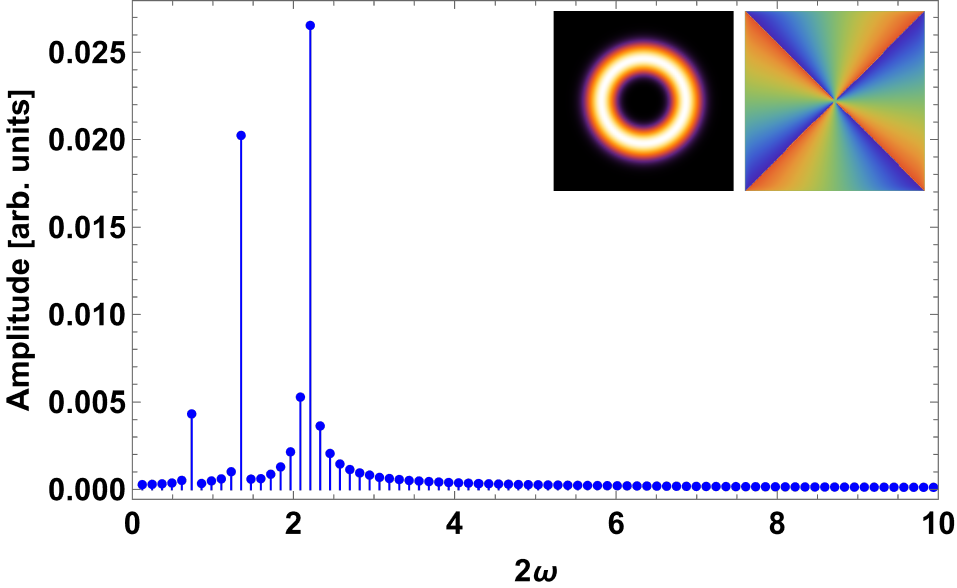}}
    \end{minipage}
     \begin{minipage}{0.33\textwidth}
        \centering
        \subfloat[]{\includegraphics[width=1.0\textwidth]{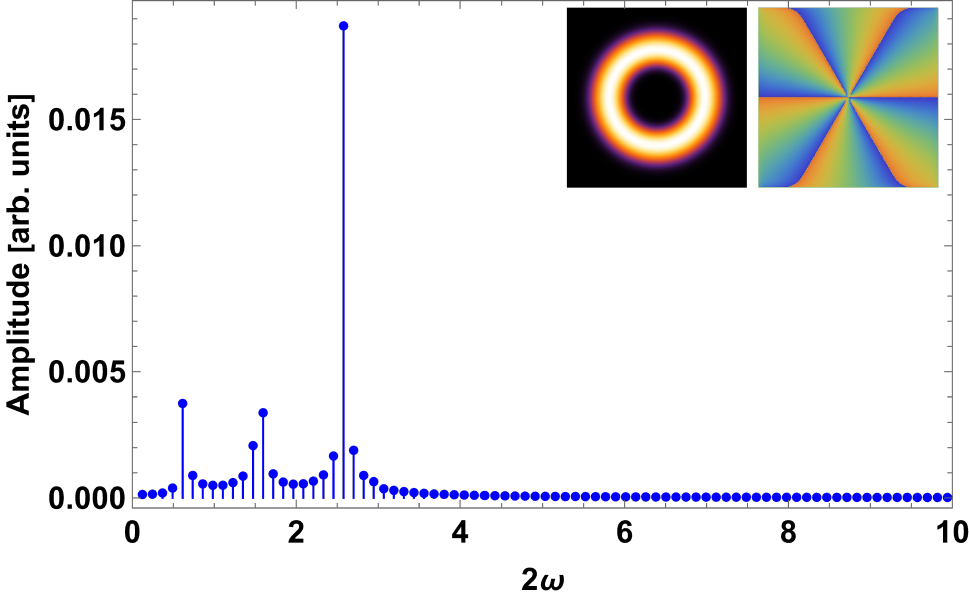}}
    \end{minipage}
        \begin{minipage}{0.33\textwidth}
        \centering
        \subfloat[]{\includegraphics[width=1.0\textwidth]{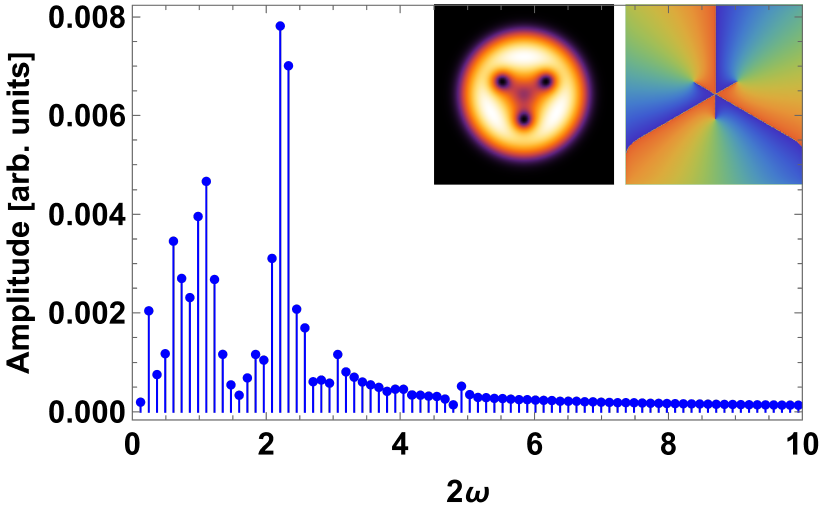}}
    \end{minipage}\hfill
    \begin{minipage}{0.33\textwidth}
        \centering
        \subfloat[]{\includegraphics[width=1.0\textwidth]{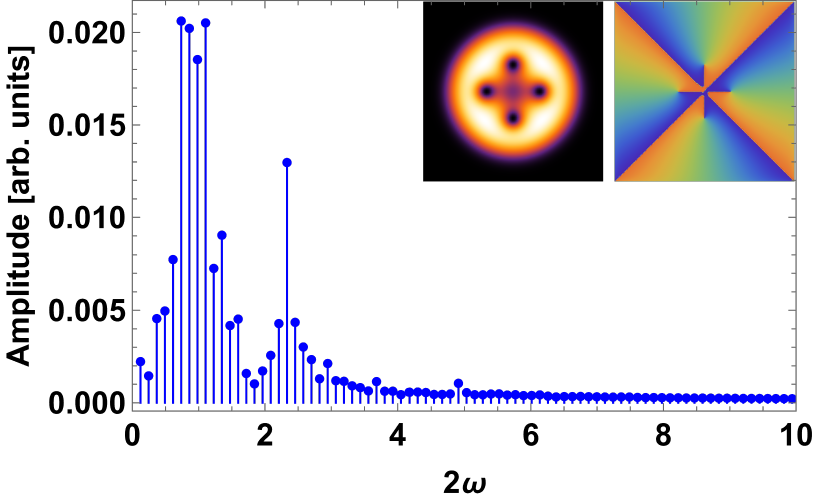}}
    \end{minipage}
     \begin{minipage}{0.33\textwidth}
        \centering
        \subfloat[]{\includegraphics[width=1.0\textwidth]{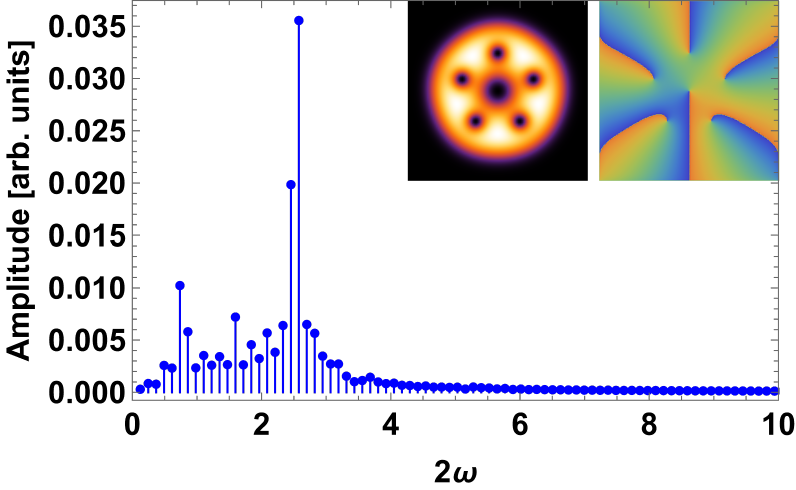}}
    \end{minipage}
         \caption{
Quadrupole mode of the shifted harmonic oscillator with $\rho_0 =2.5$. The interaction parameter is $G_{2D}=10$ for the (a-c) macrovortices and $G_{2D}=100$ for the (d-f) vortex lattices. The rotation frequencies are $\Omega=0.3$, $0.4$, and $0.5$ for panels (a) and (d), (b) and (e), (c) and (f), respectively. The insets reproduce the density and phase profiles of the ground state. We observe a small atomic density at the center of the panels (d) and (e), which could be reduced by employing larger ring radii or a lower value of the interaction parameter.
\label{fig:FFT_SHO}}
    \end{figure*}

Finally, we compare the collective modes frequency of the macrovortex and the vortex lattice states for the same value of $\Omega$. While we used $G_{2D}=10$ for all the macrovortex configurations, to have stable vortex lattices, we had to increase the value of interaction to $G_{2D} = 100$. Figure~\ref{fig:FFT_MH} contains the macrovortex and vortex lattice monopole frequency for three values of the rotation frequency $\Omega=0.3, 0.4$, and $0.5$ in the MH trap. We observe that the well-defined peak of the macrovortex case evolves to a Lorentzian-shaped one in the corresponding vortex lattice configuration. The same situation is illustrated in Fig.~\ref{fig:FFT_SHO}, where we explored the frequency spectrum of the quadrupole mode in the SHO trap. The damping of the oscillations in the lattice case, i.e., the broadening of the collective mode peak, can be attributed to the appearance of additional modes due to the oscillation of the vortices. During the collective dynamics, the separation of the vortices and, consequently, their interaction will change over time. In this process, part of their energy can be transferred to the sound wave modes, keeping the total energy of the system constant.

\section{Conclusions}
\label{sec:conclusions}

We studied the vortex configurations that appear in different ring confinements that can be implemented in bubble trap experiments. We used analytical and numerical calculations to determine the phase diagram of the system and its collective mode dynamics. We showed how the configuration of the vortices significantly affects the condensate ground state solution in the rotating frame and the frequency of the collective modes. The magnitude of the frequency of the monopole mode increases with the cloud vorticity, reflecting the lower compressibility of condensates with vortices. Also, we observed the damping of the collective modes associated with the vortex lattice formation, which can be an auxiliary method to probe the macrovortex to singly-charged vortex lattice transition in the experiments. We emphasise that the experimental conditions make the configuration of vortices hard to resolve in in-situ and free expansion images due to the fast radial expansion of the rotating cloud \cite{Guo2021}. Our calculations are expected to be valid in the small interactions limit. For current experiments, $G_{2D}$ is of the order of 100. Still, this quantity can be reduced by decreasing the number of atoms $N$, increasing the axial confinement $\tilde{Z}$, or even reducing the atomic scattering length using the Feshbach resonances.

In future works, we want to improve our variational phase diagram, including higher values for the external rotation and the atomic interaction to further map the experimental conditions. For that, we intend to use the Bogoliubov-de Gennes equations to track the dynamic instabilities of the system and then recalculate the phase diagram with less computational cost. These results will be left for future publication as they are outside the scope of the present work. We will also extend our analytical calculations to explore higher interaction regimes using the Thomas-Fermi (TF) ansatz for the ground-state solution. It will be interesting to compare the frequency evolution of the collective modes from the macrovortex to the TF-lattice regime. The latter has the vortex core and the lattice parameter (space between vortices) much smaller than the ring dimensions. While useful in many analyses, numerical methods have the disadvantage of not always elucidating the underlying physics, so the analytical expressions determined here for the frequencies of the modes are very desirable.

\begin{acknowledgments}
We thank V.S. Bagnato for the useful discussions. This work was partially supported by the S\~ao Paulo Research Foundation (FAPESP) grants 2013/07276-1 (M.A.C.), 2018/09191-7, and  	
2023/04451-9 (L.M.), and by the National Council for Scientific and Technological Development (CNPq) grants 383501/2022-9 (L.M.) and 140663/2018-5 (G.T.).
\end{acknowledgments}

\section*{Data Availability Statement}

Data is available on request from the authors.

\appendix

\section{Bubble trap potential and its approximations}
\label{app:approximations}

\begin{figure}[!htb] 
\centering
{\includegraphics[width=\linewidth]{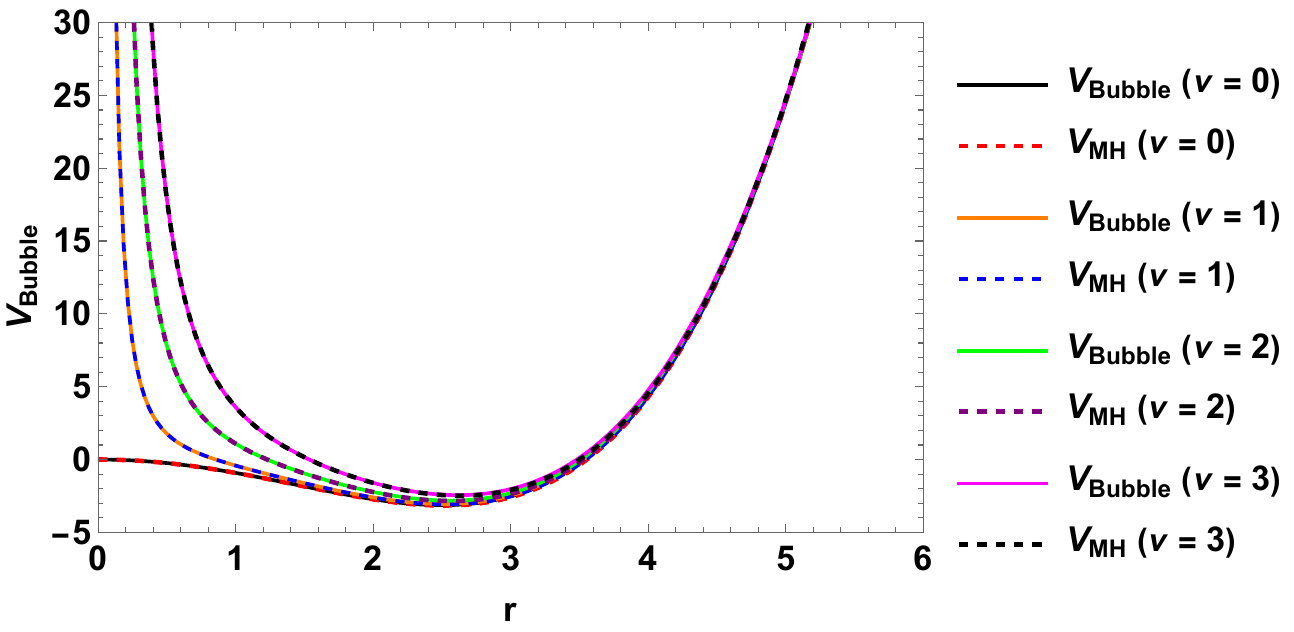}}
\caption{
Bubble trap potential and the corresponding MH approximation. The MH parameters are $\sigma = 10$ and $r_{\rm min} = 2.5$.}
\label{fig:2b}
\end{figure}

\begin{figure}[!htb] 
\centering
{\includegraphics[width=\linewidth]{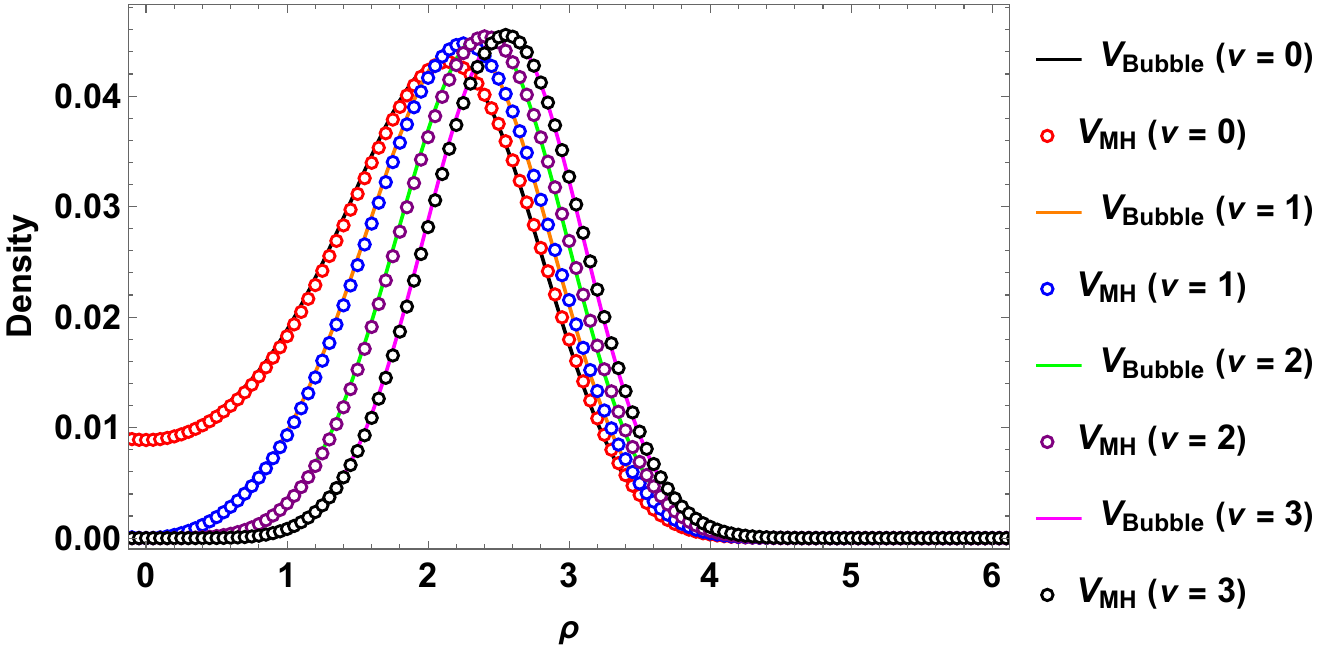}}
\caption{ 
Macrovortex density profiles obtained through numerical simulations of the imaginary-time GPE using the full expression of the ring trap and the MH approximation for $G_{2D} =10$. To produce the macrovortices, we employed rotation frequencies in the range $0.2 \leqslant \Omega \leqslant 0.4$.}
\label{fig:2}
\end{figure}

\begin{figure}[!htb] 
\centering
{\includegraphics[width=\linewidth]{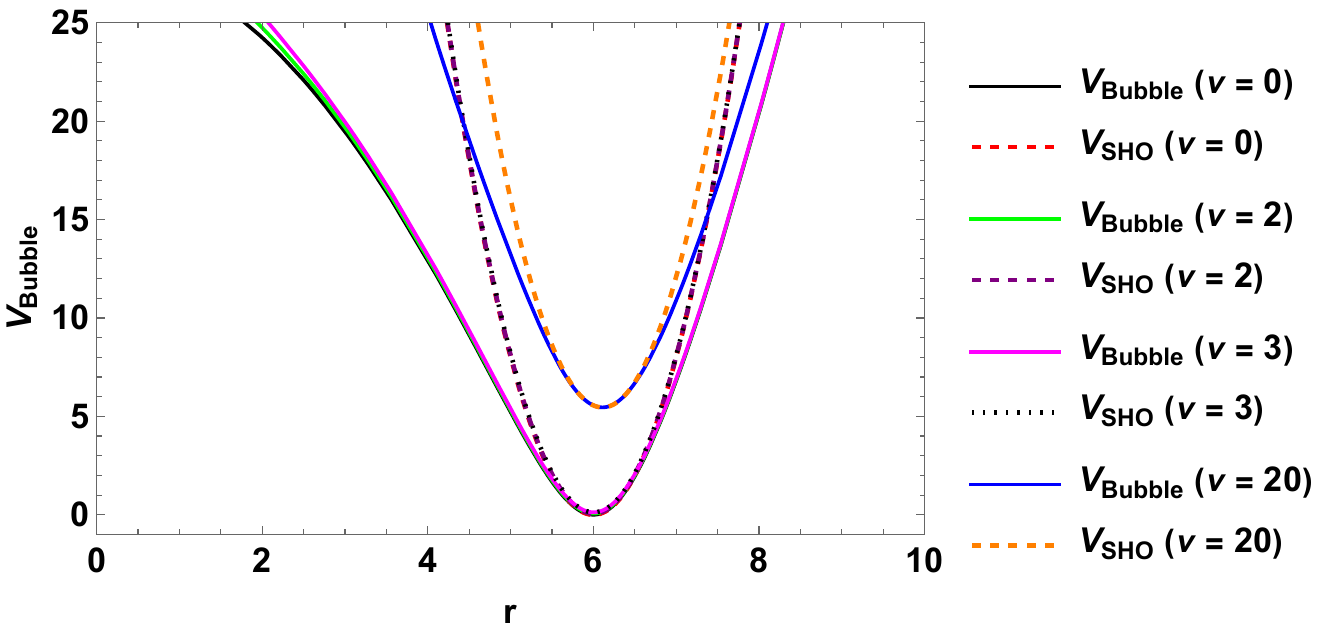}}
\caption{
Analogous to Fig.~\ref{fig:2b}, but for the SHO approximation. The SHO parameters are $\sigma = 0.25$ and $r_{\rm min} = 6$.}
\label{fig:3b}
\end{figure}

\begin{figure}[!htb] 
\centering
{\includegraphics[width=\linewidth]{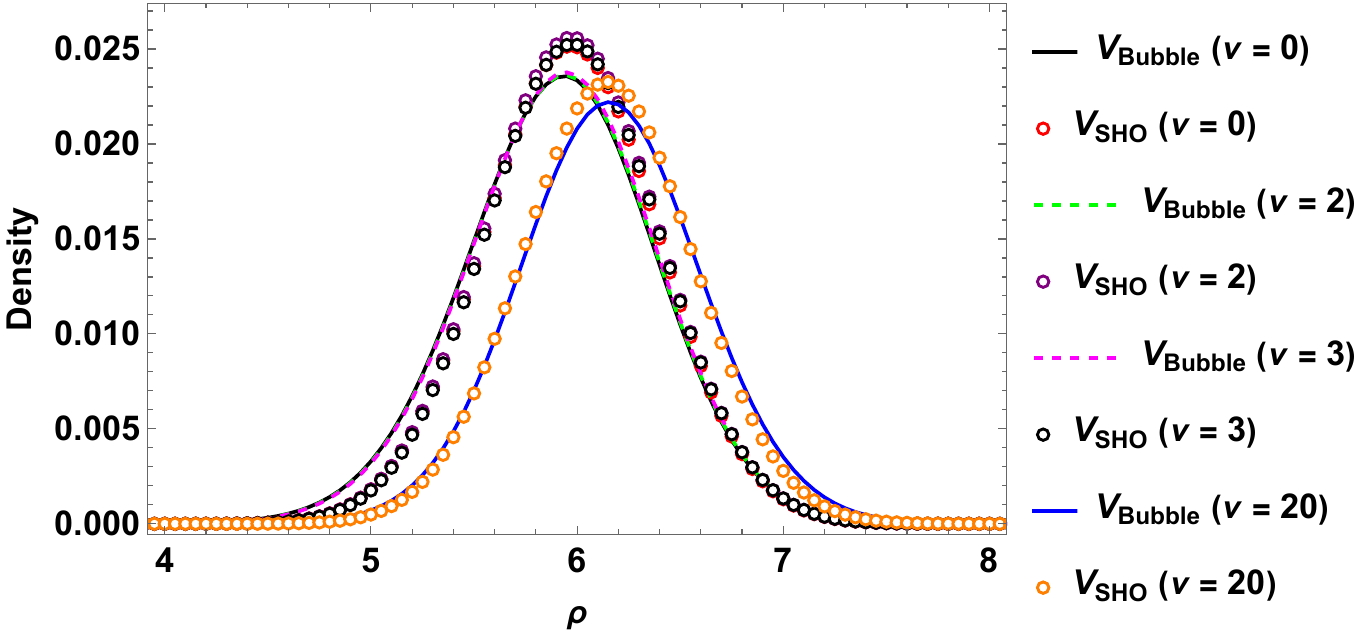}}
\caption{
Analogous to Fig.~\ref{fig:2}, but for the SHO approximation. To produce the macrovortices, we employed rotation frequencies in the range $0.1 \leqslant \Omega \leqslant 0.6$.}
\label{fig:3}
\end{figure}

To illustrate the two approximations for the ring trap, let us consider the potentials shown in Figs.~\ref{fig:2b} and ~\ref{fig:3b}. As mentioned in the main text, the MH approximation works best in the small radii and thick ring regime, while the SHO agreement improves in the opposite scenario: large radii and thin rings. For the approximations in Figs.~\ref{fig:2b} and ~\ref{fig:3b}, we compared the numerically calculated wave function of the full bubble trap with that of the approximate traps. Figures~\ref{fig:2} and \ref{fig:3} show the ground-state macrovortex solutions for the MH and SHO potentials, respectively. The quantitative agreement with the bubble trap is excellent in the case of the MH potential. We have a qualitative agreement for the SHO, which improves as the vorticity increases. 

\section{Collective modes and sum rules}
\label{sec:app_sum_rules}

Consider an arbitrary physical operator $\hat{F}$ exciting the system from its ground state $|0\rangle$ to an excited state. The response of the system ground state to the action of the operator $\hat{F}$ is characterized by the strength function, 
\begin{equation}
\label{eq:strength_function_app}
S(\omega) = \sum_{n>0} |\langle n|\hat{F}|0 \rangle|^2 \delta(\omega-\omega_n),
\end{equation} where $\omega_n = E_n-E_0$, with the Hamiltonian of the system satisfying $\hat{H} |n\rangle = E_n |n\rangle$. Determining $S(\omega)$ can be difficult since it involves the calculation of all the eigenstates of $\hat{H}$. The sum-rule technique provides a useful alternative to the explicit evaluation of $S(\omega)$. It is based on the determination of the moments $m_p$ of the strength function, 
\begin{eqnarray} \label{moments0AP}
m_p=\int_{0}^{\infty} S(\omega) \omega^p d\omega = \sum_{n>0} |\langle n|\hat{F}|0 \rangle|^2  \omega_n^p,
\end{eqnarray} which can be rewritten as mean values on the ground state of the system. For operators $\hat{F}$ such that $\langle 0| \hat{F}|0 \rangle =0$, and $p\geqslant 0$, one has
\begin{eqnarray} \nonumber \label{moments}
m_p&&= \sum_{n}  \langle 0|\hat{F}^{\dagger}|n \rangle \langle n|\hat{F}|0 \rangle  (E_n-E_0)^p \\
&&= \langle 0|\hat{F}^{\dagger}(\hat{H}-E_0)^p\hat{F}|0 \rangle,
\end{eqnarray} where we used the completeness relation $\sum_n |n\rangle \langle n | = \hat{I}$. The sum-rule approach is particularly useful when a few moments can characterize the strength function. It is straightforward to check that the moments in Eq.~(\ref{moments}) with $p>0$ can be expressed through commutators and anti-commutators involving the excitation operator and the Hamiltonian of the system $\hat{H}$ \cite{LIPPARINI1989}, as shown in Eq.~(\ref{m4minusq}).

\subsection{Excitation Operator}
\label{sec:exc_app}

We start by defining a general excitation operator using spherical coordinates and the spherical harmonics expansion. We note that these operators can be defined disregarding an overall multiplicative constant since the moments depend on ratios involving the operators both in the numerator and the denominator. The general form of the operator is 
\begin{equation}
\hat{F}(n,\ell,m)\propto \sum_{j} r_j^{2n+\ell}Y_{\ell,m}(\theta_j,\phi_j),
\end{equation} where $Y_{\ell,m}$ represents the spherical harmonics and the sum is taken over the particles of the system. The excitation with $n=1$ and $\ell=m=0$ is called the monopole mode, and the corresponding operator is
\begin{equation} \label{FM}
\hat{F}_{M}\equiv\hat{F}(1,0,0)\propto \sum_j r_j^2 = \sum_j (x_j^2+y_j^2+z_j^2).
\end{equation} 

The excitations with $n=0$ and $\ell=2$ are identified with the quadrupole modes, which are classified into five types depending on the value of $m$,
\begin{eqnarray} \label{FQ}
\hat{F}_{Q}&\equiv&\hat{F}(0,2,m)\propto\sum_j r_j^2 Y_{2,m}(\theta_j,\phi_j) \nonumber\\
&\propto&
\begin{cases}
\sum\limits_j (x_j\pm i y_j)^2,     &\text{for }  m=\pm 2, \\
\sum\limits_j (x_j\pm i y_j) z_j,   &\text{for }  m= \pm 1, \\
\sum\limits_j (x_j^2+ y_j^2-2 z_j^2), &\text{for } m= 0.
\end{cases}
\end{eqnarray}

\subsection{Quadrupole mode: two- and four-modes approaches}
\label{sec:quadrupole_app}

 We have, according to Eq.(\ref{strength}), 
\begin{eqnarray} \label{m1plus}
&&m_1^+=\sigma_0 \left(\omega_++\omega_-\right),\\
\label{m1minus}
&&m_{-1}^+=\sigma_0 \left(\frac{1}{\omega_+}+\frac{1}{\omega_-}\right)=\frac{m_1^+}{\omega_+\omega_-},\\
 \label{m2minus}
&&m_2^-=\sigma_0 \left(\omega_+^2-\omega_-^2\right)= m_1^+\left(\omega_+-\omega_-\right), \\
 \label{m3plus}
&&m_3^+=\sigma_0 \left(\omega_+^3+\omega_-^3\right)= m_1^+\left(\omega_+^2-\omega_+\omega_-+\omega_-^2\right), \\
 \label{m4minus} 
&&m_4^-=\sigma_0 \left(\omega_+^4-\omega_-^4\right)= m_2^-\left(\omega_+^2+\omega_-^2\right). 
\end{eqnarray}
Using Eqs.~(\ref{m1plus}) and (\ref{m2minus}) yields
\begin{eqnarray} \nonumber \label{eq:Ewithm-1}
\omega_+-\omega_-=\frac{m_2^-}{m_1^+}, \\
\omega_+\omega_-=\frac{m_1^+}{m_{-1}^+},
\end{eqnarray}
and solving for the energies,
\begin{equation} \label{modefreq}
\omega_{\pm}=\frac{1}{2}\left[\sqrt{\left(\frac{m_2^-}{m_1^+}\right)^2+4\frac{m_1^+}{m_{-1}^+} }\mp \frac{m_2^-}{m_1^+}\right].
\end{equation}

Equations~(\ref{m3plus}) and (\ref{m4minus}) provide the identity
\begin{equation}
\label{eq:identity_mminus1}
\frac{m_3^+}{m_1^+} = \frac{m_4^-}{m_2^-}-\frac{m_1^+}{m_{-1}^+}.
\end{equation}
Finally, to avoid the calculation of the $m_{-1}$ moments which are intricate, we rewrite Eq.~({\ref{eq:Ewithm-1}}) using the identity above, which yields Eq.~(\ref{twomodes}) in the main text.

A full treatment of the 4-mode system would be very complicated. Instead, we assume that two doubly-degenerate energy levels are present in our analysis. Then, in the sum-rule calculation, we consider Eq.~(\ref{strength}) with
\begin{eqnarray}
S_+(\omega) &=& \sigma_H \delta(\omega - \omega_H) + \sigma_L  \delta(\omega - \omega_L),\nonumber\\
S_-(\omega) &=& \sigma_H^{\dagger} \delta(\omega - \omega_H) + \sigma_L^{\dagger}   \delta(\omega - \omega_L).
\end{eqnarray}
That is, we assume the degenerescence between the energies of the higher ($\sigma_H$ and $\sigma_H^{\dagger}$) and lower ($\sigma_L$ and $\sigma_L^{\dagger}$) branches of the $m = \pm 2$ modes~\cite{Cozzini2006b}.

This gives us six equations,
\begin{eqnarray}
\label{todas1} &&m_0^-=\sigma_H+\sigma_L-\sigma_H^{\dagger}-\sigma_L^{\dagger}=0, \\
\label{todas2} &&m_{-1}^+-=\frac{\sigma_H}{\omega_H}+\frac{\sigma_L}{\omega_L}+\frac{\sigma_H^{\dagger}}{\omega_H}+\frac{\sigma_L^{\dagger}}{\omega_L}, \\
\label{todas3} &&m_1^+=\sigma_H\omega_H+\sigma_L\omega_L+\sigma_H^{\dagger}\omega_H+\sigma_L^{\dagger}\omega_L, \\
\label{todas4} &&m_2^-=\sigma_H\omega_H^2+\sigma_L\omega_L^2-\sigma_H^{\dagger}\omega_H^2-\sigma_L^{\dagger}\omega_L^2, \\
\label{todas5} &&m_3^+=\sigma_H\omega_H^3+\sigma_L\omega_L^3+\sigma_H^{\dagger}\omega_H^3+\sigma_L^{\dagger}\omega_L^3, \\
\label{todas6} &&m_4^-=\sigma_H\omega_H^4+\sigma_L\omega_L^4-\sigma_H^{\dagger}\omega_H^4-\sigma_L^{\dagger}\omega_L^4,
\end{eqnarray}
to be solved for six variables.
We start with Eqs.~(\ref{todas1})-(\ref{todas4}) and solve for $\sigma_{H},\sigma_{H}^{\dagger}$, $\sigma_{L}$, and $\sigma_{L}^{\dagger}$. Then, we replace these solutions in Eqs.~(\ref{todas5}) and (\ref{todas6}), which yields
\begin{eqnarray}  \nonumber
&&\frac{m_4^-}{m_2^-}=\omega_H^2+\omega_L^2, \\
&&\frac{m_3^+}{m_1^+}=\frac{m_4^-}{m_2^-}-\omega_H^2\omega_L^2 \frac{m_{-1}^+}{m_1^+}.
\end{eqnarray}
Then, we find that 
\begin{equation}
\omega_{H,L}^4-\left(\frac{m_4^-}{m_2^-}\right)\omega_{H,L}^2+\frac{m_{1}^+}{m_{-1}^+}\left(\frac{m_4^-}{m_2^-}-\frac{m_3^+}{m_1^+}\right) =0,
\end{equation}
with the solutions presented in Eq.~(\ref{fourmodes}) of the main text.

\subsection{Details of the moments calculations}
\label{ap:moments_calculations}

The moments are given by   
\begin{flalign}
\nonumber
&m_1^+= \langle 0 | [F_-,[H,F_+]]|0 \rangle, \\
 \nonumber
&m_2^-=  \langle 0 |[ [F_-,H],[H,F_+] ]|0 \rangle, \\
 \nonumber
&m_3^+= \langle 0 |[ [F_-,H],[H,[H,F_+]] ]|0 \rangle, \\ 
\label{m4minusq} 
&m_4^-= \langle 0 |[ [[F_-,H],H],[H,[H,F_+]] ]|0 \rangle.
\end{flalign}

For the Mexican hat trap, we obtain
\begin{eqnarray} \nonumber
&&\frac{m_2^-}{m_1^+} = \left[\frac{1}{{\Omega}}\frac{E_{rot}}{E_{ho}}- 4  {\Omega}\right], \\  \nonumber
&&\frac{m_3^+}{m_1^+} = 2  \left[\frac{E_{kin}}{E_{ho}}-1+6{\Omega}^2-3\frac{E_{rot}}{E_{ho}}+3\lambda\frac{\langle \rho^4\rangle}{\langle \rho^2\rangle}\right], \\ \nonumber
&&\frac{m_{4}^{-}}{m_{2}^{-}}  =  4 \bigg\{\frac{\frac{E_{kin}}{E_{ho}}-1+2{\Omega^{2}}-\frac{3}{2}\frac{E_{rot}}{E_{ho}}}{1-\frac{1}{4{\Omega}^{2}}\frac{E_{rot}}{E_{ho}}} \nonumber \\
 && \qquad \qquad +\frac{\frac{1}{4{\Omega}^{2}}\frac{E_{rot}}{E_{ho}}-3\lambda\left(\frac{1}{2{\Omega}}\frac{\left\langle \rho^{2}{L}_{z}\right\rangle }{\left\langle {\rho}^{2}\right\rangle }-\frac{\left\langle \rho^{4}\right\rangle }{\left\langle \rho^{2}\right\rangle }\right)}{1-\frac{1}{4{\Omega}^{2}}\frac{E_{rot}}{E_{ho}}} \bigg\}.
\end{eqnarray} 

For the SHO trap, we have
\begin{eqnarray}
&&\frac{m_2^-}{m_1^+} = \left[\frac{1}{{\Omega}}\frac{E_{rot}}{E_{ho}}- 4  {\Omega}\right], \\ \nonumber
&&\frac{m_3^+}{m_1^+} = 2 \left[\frac{E_{kin}}{E_{ho}}+1+6{\Omega}^2-3\frac{E_{rot}}{E_{ho}}-\frac{3}{4} \rho_0\frac{\langle \rho \rangle}{\langle \rho^2\rangle}\right], \\ 
&&\frac{m_{4}^{-}}{m_{2}^{-}}  = 4 \bigg\{\frac{\frac{E_{kin}}{E_{ho}}+1+2{\Omega^{2}}-\frac{3}{2}\frac{E_{rot}}{E_{ho}}}{1-\frac{1}{4{\Omega}^{2}}\frac{E_{rot}}{E_{ho}}} \nonumber \\
&& \qquad \qquad -\frac{\frac{1}{4{\Omega}^{2}}\frac{E_{rot}}{E_{ho}}-\frac{3}{4}\rho_{0}\left(\frac{1}{2{\Omega}}\frac{\left\langle {L}_{z}/\rho\right\rangle }{\left\langle {\rho}^{2}\right\rangle }-\frac{\left\langle \rho\right\rangle }{\left\langle \rho^{2}\right\rangle }\right)}{1-\frac{1}{4{\Omega}^{2}}\frac{E_{rot}}{E_{ho}}}\bigg\}.
\end{eqnarray}

\noindent In the equations above, we have the rotational energy,
\begin{equation}
E_{rot}=\Omega \langle L_z \rangle,
\end{equation} with the angular momentum given by 
\begin{equation}
L_z=\left( x \, p_y - p_x \, y \right).
\end{equation}  We also used the identity
\begin{equation}
\frac{\langle L_z \rangle}{2\Omega\langle \rho^2 \rangle}=\frac{1}{4\Omega^2}\frac{E_{rot}}{E_{ho}}.
\end{equation} 

\subsection{Virial Theorem}
\label{sec:ap1}

In the following discussions, we restrict ourselves to a situation where $N$ particles are confined in a Mexican hat or a shifted harmonic potential and experience contact interactions described by a delta function. The Hamiltonian is given by Eq.~(\ref{rotFrame}), and we assume that the state of the system is stationary. We define the expectation value of a quantity $x$ as the usual inner product with the wave function, i.e., $\langle x\rangle \equiv \int d^2r \; \psi^*x\psi$. Then, the expectation value of $\sum_i \bm{q}_i \cdot\bm{p}_i$, with ${\bf{q}}_i$ corresponding to the generalized coordinates of the $i$-th particle, is constant in time,
\begin{equation}
\frac{d}{dt}\left\langle \sum_i \bm{q}_i \cdot\bm{p}_i   \right\rangle =0.
\end{equation}
On the other hand, Heisenberg's equation of motion gives
\begin{equation}
\frac{d}{dt} \sum_i \bm{q}_i\cdot \bm{p}_i = \frac{i}{\hbar} \left[\hat{H},\sum_i \bm{q}_i\cdot \bm{p}_i\right].
\end{equation}

We derived the Virial relation for two different trap potentials: the Mexican hat, Eq.~(\ref{eq:MHpotential}), and the SHO, Eq.~(\ref{eq:SHOPotential}). For the MH trap,
\begin{equation} \label{virialMH1}
  E_{kin} +  E_{int} +  E_{ho} - \lambda\langle \rho^4 \rangle =0,
\end{equation}
which can be rewritten as
\begin{equation}
2 E_{kin}  -2  E_{trap}^{MH}  + 2  E_{int}  -\lambda\langle \rho^4 \rangle =0,
\end{equation} with
\begin{equation}
 E_{trap}^{MH} = -\frac{1}{2}\langle \rho^{2}\rangle +\frac{\lambda}{2}\langle \rho^{4}\rangle. 
\end{equation} On the other hand, for the SHO, \begin{equation} 
2E_{kin}  +2 E_{int}  -2  E_{ho}  + \rho_0 \langle \rho \rangle =0,
\end{equation} which can be rewritten as
\begin{equation} 
2 E_{kin}  -2  E_{trap}^{\rm SHO} + 2  E_{int}  - \rho_0 \langle \rho \rangle +\rho_0^2=0,
\end{equation} with \begin{equation}
 E_{trap}^{\rm SHO} = \frac{1}{2}\langle (\rho-\rho_0)^{2}\rangle. 
\end{equation}

\section{Compressibility Sum Rules}
\label{compressibility}

More accurate results for the sum rule frequencies involve determining the compressibility moment $m_{-1}$. In the case of the quadrupole excitation in the two-mode approach, we were able to avoid the explicit calculation of $m_{-1}$ using the identity of Eq.~(\ref{eq:identity_mminus1}). A similar substitution is not possible for the case of the four-mode approximation, and the moment $m_{-1}$ cannot be determined directly from the numerical wave function. Hence, we assumed a Thomas-Fermi profile to be able to perform the calculation. The closer the system is to the TF limit, the better this approximation will be.

Differently from $m_1$ and $m_3$, it is more convenient to relate the moment $m_{-1}$ to the static response function \cite{Pitaevskii2016}. A rule for calculating $m_{-1}$ is through the explicit determination of the polarization.

\subsection{Monopole}

To calculate the ratio $m_1/m_{-1}$, which involves the inverse energy weighted moment, we considered 
\begin{equation}
2 m_{-1}=-\chi_M(0),
\end{equation} where the susceptibility $\chi_M(0)$ characterizes the static limit of the dynamic response function.
For the moment $m_1$, we have the expression
\begin{equation}
m_1=\frac{2N\hbar^2\langle{\tilde{\rho}}^2\rangle}{\tilde{M}},
\end{equation} where $\tilde{M}$ corresponds to the mass of the particle and the mean value of the position squared is taken with respect to the ground state of the system.

To determine $\chi_M(0)$, first we considered the Hamiltonian of Eq.~(\ref{rotFrame}) with the perturbation
\begin{equation}
H_{pert} =\frac{1}{2}\tilde{M}\tilde{\omega}_0^2\gamma\sum_i \tilde{\rho}_i^2. 
\end{equation}
After, we applied the procedure to produce dimensionless quantities, 
\begin{eqnarray}\nonumber
\tilde{\rho}&=&\tilde{\ell}_{\gamma} r_1,\\ \nonumber
\tilde{\mu} &=& \mu_1 \hbar \tilde{\omega}_{\gamma},\\ \nonumber
\tilde{\psi}&=&\sqrt{\frac{N}{\sqrt{2\pi}{Z}\tilde{\ell}_{\gamma}^2}}\psi_1,
\end{eqnarray} with  $\tilde{\omega}_{\gamma}=(1+\gamma)^{1/2}\tilde{\omega}_0$ and $\tilde{\ell}_{\gamma}=\sqrt{\hbar/\tilde{M} \tilde{\omega}_{\gamma}}$.

We solved the corresponding time-independent GPE, obtained substituting $\psi_1(t)=\psi_1(0) e^{-\mu_1 t}$ in Eq.(\ref{eq:GPE}). The term $ \bm{\Omega}\cdot\bm{L}$ was replaced by $\bm{\Omega}\cdot\bm{r} \times\bm{v}$, where we considered the solid-body rotation $\bm{v}=\bm{\Omega}\times\bm{r}$. Using the TF approximation for the MH trap potential yields 
\begin{equation}
n_{vl}^{MH}(r_1)=|\psi_1|^2=\frac{\mu_1+\frac{1}{2}\left[\Omega_1^2 r_1^2+r_1^2-\lambda_1(\gamma)r_1^4\right]}{G_{2D}},
\end{equation} where 
\begin{equation}
\lambda_1(\gamma) = \frac{\lambda\tilde{\ell}_{\gamma}^2}{(1+\gamma)},
\end{equation} and $\Omega_1=\tilde{\Omega}/\tilde{\omega}_{\gamma}$.
For $\chi_M(0)$, we have the identity 
\begin{equation}\label{chiM}
\chi_M(0)=\frac{2N}{M}\frac{\partial\langle \tilde{\rho}^2\rangle}{\partial\tilde{\omega}_{\gamma}^2}\bigg|_{\gamma\rightarrow 0}=\frac{2N}{M}\frac{\partial\gamma}{\partial\tilde{\omega}_{\gamma}^2}\frac{\partial\langle \tilde{\rho}^2\rangle}{\partial\gamma}\bigg|_{\gamma\rightarrow 0}.
\end{equation} Then, 
\begin{equation}
\frac{\partial\langle \tilde{\rho}^2\rangle}{\partial\gamma}=\frac{\partial[\tilde{\ell}_{\gamma}^2\langle r_1^2\rangle]}{\partial\gamma}=2\tilde{\ell}_{\gamma}\frac{\partial\tilde{\ell}_{\gamma}}{\partial\gamma}\langle r_1^2\rangle+ \tilde{\ell}_{\gamma}^2\frac{\partial \langle r_1^2\rangle}{\partial\gamma}.
\end{equation} Angular momentum conservation gives
\begin{equation}
\frac{\partial}{\partial\gamma}[\Omega_1\langle r_1^2\rangle]=0.
\end{equation} Remembering that 
\begin{equation}
\langle r_1^2 \rangle = \frac{(1+\Omega_1^2)}{2\lambda_1},
\end{equation} we get
\begin{equation}
\frac{m_1}{m_{-1}}=-2\hbar^2\tilde{\omega}_0^2\frac{\langle r_1^2\rangle \tilde{\ell}_{\gamma}^2}{2\tilde{\ell}_{\gamma}\frac{\partial\tilde{\ell}_{\gamma}}{\partial\gamma}\langle r_1^2\rangle+  \frac{\tilde{\ell}_{\gamma}^2\frac{\partial\lambda_1}{\partial\gamma}\frac{\partial\langle r_1^2\rangle}{\partial\lambda_1}}{1+\frac{\Omega_1}{\langle r_1^2\rangle}\frac{\partial\langle r_1^2\rangle}{\partial\Omega_1}}}.
\end{equation}
Then, taking the limit $\gamma \rightarrow 0$, yields
\begin{equation}
 \sqrt{\frac{m_1}{m_{-1}}}=\hbar\tilde{\omega}_0 \sqrt{6\Omega^2+2},
\end{equation} with $\Omega= \tilde{\Omega}/\tilde{\omega}_0$

Analogously for the SHO trap, with a TF density profile given by
\begin{equation}
n^{\rm SHO}_{vl}(r_1) = \frac{\mu_1+\frac{1}{2}\left[\Omega_1^2r_1^2-r_1^2-\frac{1}{1+\gamma}({r}_0^2-2r_1{r}_0)\right]}{G_{2D}},
\end{equation} where ${r}_0 = \tilde{\rho}_0/\tilde{\ell}_{\gamma}$. We find
\begin{equation} \label{eqredesho}
\sqrt{\frac{m_1}{m_{-1}}}=\hbar\tilde{\omega}_0\sqrt{\frac{12{\eta\prime}^2(1-{\Omega}^2)^3+80 r_0^2(1+3{\Omega}^2)}{20r_0^2(5-{\Omega}^2)-{\eta\prime}^2(1-{\Omega}^2)^3}}.
\end{equation} where
\begin{equation} \label{dsho}
\eta\prime=\left(\frac{6}{\pi}\frac{G_{2D}}{r_0}\right)^{1/3}.
\end{equation}

\subsection{Quadrupole}

Using the same dimensionless quantities of the main text, we calculated the static quadrupole response through the perturbation with strength $\varepsilon$,
\begin{equation}
\delta V_{ext} = \frac{1}{2}\varepsilon(x^2-y^2),
\end{equation} which introduces an asymmetry in the confining potential. 
More specifically, we determined the compressibility mode by calculating the limit
\begin{equation} \label{m1minusAppendice}
m_{-1}^+ = {4 N}\lim_{\varepsilon \rightarrow 0}\langle x^2+y^2 \rangle \left(\frac{\partial\delta_{\varepsilon}}{\partial\varepsilon} \right),
\end{equation} where
\begin{equation}
\delta_{\varepsilon}=\frac{\langle y^2-x^2\rangle}{\langle x^2+y^2\rangle}.
\end{equation}  First, we considered the perturbed density profile for the MH potential,
\begin{align}
n_{vl}^{MH}\left(\rho\right) & =\frac{\mu+\frac{{\Omega}^{2}}{2}\left(x^{2}+y^{2}\right)+\frac{1}{2}\left[\left(1+\varepsilon\right)x^{2}+\left(1-\varepsilon\right)y^{2}\right]}{G_{2D}} \nonumber\\
 & -\frac{\frac{\lambda}{2}\left(x^{2}+y^{2}\right)^{2}}{G_{2D}}.
\end{align}
By using Eq.~(\ref{m1minusAppendice}), it is straightforward to show that
\begin{equation}
m_{-1}^+ = \frac{\pi N}{3G_{2D}}(R_2^6-R_1^6),
\end{equation} where the boundaries of the TF profile are $R_1$ and $R_2$, with $R_2>R_1$.
Using the simplified notation $R_{\pm}^2=R_2^2\pm R_1^2$, we get
\begin{equation}
\frac{m_1^+}{m_{-1}^+}=4\lambda \frac{R_+^2R_-^4}{3R_+^4+R_-^4},
\end{equation} where we used that 
\begin{equation}
m_{1}^+ = 8N\langle \rho^2\rangle.
\end{equation}  Then, we applied the TF profile to determine the MH solution,
\begin{equation}
\sqrt{\frac{m_1^+}{m_{-1}^{+}}}=2\sqrt{\frac{(1+{\Omega}^2)\eta^2}{3(1+{\Omega}^2)^2+\eta^2} },
\end{equation} with 
\begin{equation} 
 \eta = \left(\frac{12 G_{2D}\lambda^2}{\pi} \right)^{1/3}.
\end{equation}
On the other hand, for the macrovortex profile with charge $\nu$, we have
\begin{equation}
 \sqrt{\frac{m_1^+}{m_{-1}^{+}}}=2 {R_-^2} \sqrt{\frac{12\nu^2+R_+^4-R_-^4}{3R_+^8-2R_-^4R_+^4-R_-^8} }.
\end{equation}

For the SHO case, considering the simplified notation of the ring mean radius $R= ({R_1+R_2})/{2}$ and width $d=R_2-R_1$, we get
\begin{equation}
 \sqrt{\frac{m_1^+}{m_{-1}^{+}}}=\sqrt{\frac{8 (1-{\Omega}^2)(3d^4+20R^2d^2)}{5(3d^4+40d^2R^2+48R^4)}}.
\end{equation}
Whereas, if we consider the macrovortex profile,
\begin{widetext}
\begin{equation} \label{eqfinal}
 \sqrt{\frac{m_1^+}{m_{-1}^{+}}}=\sqrt{\frac{8 d^2\left[(d^2-4R^2)^2(3d^2+20R^2)-48(d^2-20R^2)\nu^2\right]}{5(d^2-4R^2)^2(3d^4+40d^2R^2+48R^4)}}.
\end{equation}
\end{widetext}

\bibliography{references}

\end{document}